\documentclass[fp,twocolumn]{jpsj3}

\usepackage{txfonts}
\usepackage{bm}
\usepackage{braket}
\usepackage{color}
\usepackage{cite}
\usepackage{ulem}

\title{Atomic-Scale Magnetic Toroidal Dipole under Odd-Parity Hybridization}

\author{Megumi Yatsushiro and Satoru Hayami}
\inst{Department of Physics, Hokkaido University, Sapporo 060-0810, Japan} 

\abst{
Magnetic toroidal dipole (MTD) is one of a fundamental constituent to induce magneto-electric effects in the absence of both spatial inversion and time-reversal symmetries.
We report on a microscopic investigation of the atomic-scale MTD in solids by taking into account the orbital degree of freedom with a different parity.
We construct an effective two-orbital $d$-$f$ tight-binding model on a polar tetragonal system for describing the atomic-scale MTD, which are obtained by incorporating the atomic spin-orbit coupling and odd-parity hybridization.
The effective model exhibits two types of the MTDs: 
in-plane $x, y$ components 
activated through spontaneous ferromagnetic ordering or external magnetic field
and an out-of-plane $z$ component by a spontaneous odd-parity hybridization without spin moments.
We show that the intra-orbital (inter-orbital) Coulomb interaction in multi-orbital systems plays an important role in stabilizing the in-plane (out-of-plane)
MTD orderings. 
We also examine the magneto-electric effect under each MTD ordering by calculating a linear response tensor.
We show that the odd-parity hybridization enhances the magneto-electric effect 
for the in-plane MTDs, while it suppresses that for the out-of-plane MTD.
}

\begin{document}
\maketitle

\section{Introduction}
Magneto-electric (ME) effect is an intriguing phenomenon where an external electric (magnetic) field induces magnetization (electric polarization) in the systems without both spatial inversion and time-reversal symmetries~\cite{curie1894symetrie, spaldin2005renaissance, khomskii2009trend}.
There are several microscopic origins for such a coupling between electric and magnetic degrees of freedom.  
One of the most prominent examples is the ferroelectricity under the spin orderings which break the spatial inversion symmetry, as seen in a spiral magnetic structure~\cite{katsura2005spin, mostovoy2006ferro, xiang2008spin, malashevich2008first, arima2007ferroelectricity, tokura2014multiferroics}. 
Another example is found in a collinear magnetic structure with concomitant charge disproportionation in the presence of magnetostriction~\cite{van2008multiferroicity}.
These ME couplings driven by electronic degrees of freedom might give rise to gigantic ME responses, which are expected for potential applications to next-generation spintronics devices.

\begin{figure}[h!]
\centering
\includegraphics[width=85mm]{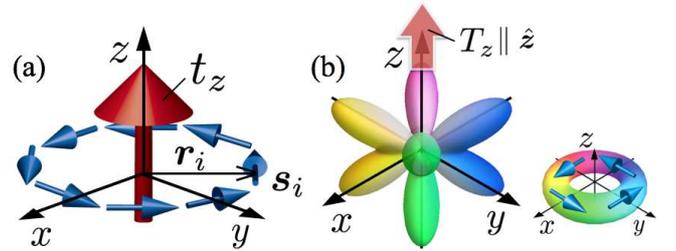}
\caption{
(Color online) Schematic pictures of (a) the MTD from a vortex-type spin cluster and (b) the MTD from an odd-parity $d$-$f$ orbital hybridization breaking time-reversal symmetry.
 In (b), the shape represents the atomic-scale charge distribution and the color represents 
the direction of the $xy$-plane orbital angular momentum shown in the lower-right panel. \label{fig:toroidal}
}
\end{figure}

A magnetic toroidal dipole (MTD) is one of fundamental objects showing the ME effect, which has been expressed as 
\begin{align}
\label{eq:MTD} 
{\bm t} \propto
\sum_i {\bm r}_i \times {\bm S}_i,
\end{align}
where ${\bm r}_i$ is the position vector and $\bm{S}_i$ is the localized spin~\cite{ederer2007towards, spaldin2008toroidal}.
The expression in Eq.~(\ref{eq:MTD}) indicates that the MTD is induced for nonzero vector products of 
$\bm{r}_i$ and $\bm{S}_i$, as schematically shown in Fig.~\ref{fig:toroidal}(a).
Such an MTD has been found in various magnetic insulators, 
such as ${\rm Cr}_2 {\rm O}_3$ under a strong magnetic field~\cite{popov1999magnetic},
${\rm Ga}_{2-x}{\rm Fe}_x{\rm O}_3$ by the X-ray scattering measurement~\cite{arima2005resonant},
${\rm LiCoPO}_4$ by the second harmonic generation measurement~\cite{van2007observation,zimmermann2014ferroic}, and 
${\rm LiFeSi}_2{\rm O}_6$ by the ME measurement~\cite{toledano2015primary}.

Recently, a concept of the MTD has been extended to various situations beyond magnetic insulators.
For example, the MTD ordered state has been indicated in a metallic compound ${\rm UNi}_4 {\rm B}$~\cite{mentink1994magnetic,hayami2014toroidal} by detecting the current-induced magnetization~\cite{saito2018evidence}. 
Meanwhile, theoretical studies show extended expressions of the MTD by considering other electronic degrees of freedom, such as the kinetic orbital moment~\cite{gao2018microscopic} and the atomic orbital angular momentum~\cite{hayami2018microscopic}.

Among them, we focus on the atomic-scale MTD, which is induced by the parity-mixing hybridization, 
such as $p$-$d$ and $d$-$f$ orbitals, without the time-reversal symmetry.
When taking into account the parity-mixing hybridization, the MTD degree of freedom is described by a vortex-type structure of the orbital angular momentum at the atomic site, as schematically shown in Fig.~\ref{fig:toroidal}(b), which is qualitatively different from the spin-cluster MTD in Eq.~(\ref{eq:MTD}) [Fig.~\ref{fig:toroidal}(a)].
In other words, both the orbital and spin degrees of freedom play an important role in the atomic-scale MTD, while the spin degree of freedom is dominant for the spin-cluster MTD. 
Reflecting such a difference, the atomic-scale MTD might be realized in the systems with a strong parity-mixing hybridization and lead to anomalous ME effects and transport properties through orbital fluctuations. 
However, such an atomic-scale MTD has not been identified in experiments yet. 
Toward experimental observations, it is desirable to clarify when and how such an MTD is stabilized 
at the microscopic level. 
It is also informative to examine the nature of physical phenomena, such as the ME effects, under the MTD ordering.

In the present study, we investigate the nature of the atomic-scale MTD in a noncentrosymmetric crystal from a microscopic point of view.  
Starting from a multi-orbital model including five $d$ and seven $f$ orbitals in a polar tetragonal crystal, we construct a minimal two-orbital model to examine the atomic-scale MTD degree of freedom.
By carefully classifying the electronic degrees of freedom in terms of multipoles, we show that two types of the MTDs are potential order parameters.
The one is the in-plane $x, y$ components of the MTDs composed of the spin and orbital degrees of freedom, while the other is the out-of-plane $z$ component composed of only the orbital degree of freedom. 
We show that the interplay among the spin-orbit coupling (SOC), the odd-parity hybridization, and the Coulomb interaction plays an important role in stabilizing the MTD orderings.  
Furthermore, we examine the ME responses for the atomic-scale MTD orderings by using the linear response theory.
We find that the odd-parity hybridization enhances the ME effect for the in-plane MTD, whereas it suppresses the ME effect for the out-of-plane MTD.

The organization of this paper is as follows.
In Sect.~\ref{sec:model}, we derive an effective two-orbital model including the atomic-scale MTD degree of freedom in a polar tetragonal crystal. 
We show that two types of the MTD can be activated in the low-energy model.
In Sect.~\ref{sec:stability}, we discuss the stability of these MTDs from the energetics viewpoint.
In Sect.~\ref{sec:ME}, we investigate the ME effect under the two types of the MTD orderings.
Section~\ref{sec:summary} is devoted to summary.
In Appendix~\ref{ap:real_description}, the definitions of the $d$ and $f$ orbital wave functions and multipole expressions are shown.
In Appendix~\ref{ap:ASOI}, we discuss the antisymmetric spin-orbit interactions between the orbitals with the same and different parities, respectively.

\section{Model with Parity-Mixing Orbitals}
\label{sec:model}
The MTD as the atomic-scale degree of freedom can be activated in multi-orbital systems with the different parities~\cite{hayami2018microscopic, hayami2018classification}.
In this section, we construct an effective model including the atomic-scale MTD degree of freedom in a $d$-$f$ orbital system. 
In Sect.~\ref{subsec:CEF}, we introduce the atomic base in a local Hamiltonian under the crystalline electric field (CEF) and the atomic SOC.
In Sect.~\ref{subsec:multipole}, we describe active multipoles in the two-orbital $d$-$f$ base.
 In Sect.~\ref{subsec:dfmodel}, we show the tight-binding model and introduce the atomic MTD moments.  
 
\begin{figure*}[t]
\centering
\includegraphics[width=176mm]{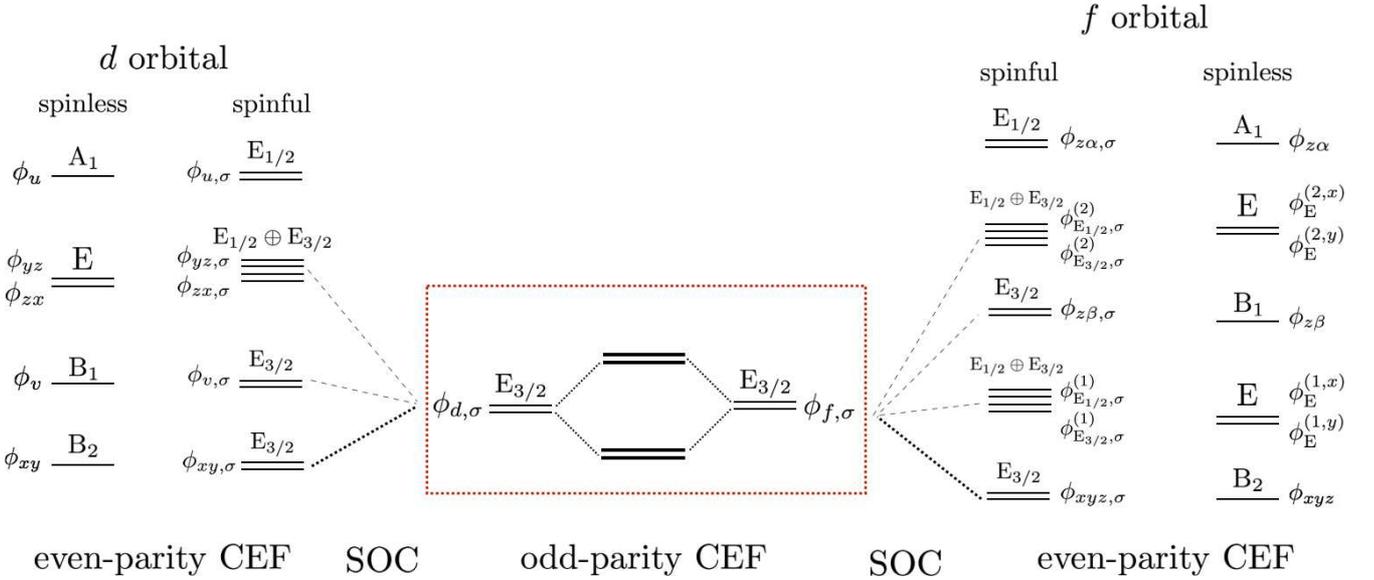}
\caption{
(Color online) Schematic energy levels under the local Hamiltonian in Eqs.~(\ref{eq:CEF}) and (\ref{eq:Hls}).
The energy level splittings for $d$ and $f$ orbitals under the even-parity CEF and the SOC are shown in the left and right panels, respectively.
The CEF parameters in Eq.~(\ref{eq:CEFeven}) are $B_{20}=1.0$, $B_{40}=0.8$, $B_{44}=0.6$, $B_{60}=0.4$, and $B_{64}=0.2$.  
The atomic wave functions and their irreducible representations in the point group $C_{\rm 4v}$ are also shown. 
The middle of the panel is the energy splitting under the odd-parity CEF in Eq.~(\ref{eq:CEFodd}) due to the parity-mixing hybridization.
The dashed square represents the low-energy levels considered in the present study.
\label{fig:CEF_split}}
\end{figure*}

\subsection{Atomic basis functions}
\label{subsec:CEF}
In order to derive a minimal model including the atomic MTD degree of freedom, we start from general five $d$ orbitals, ($\phi_u$, $\phi_v$, $\phi_{yz}$, $\phi_{zx}$, $\phi_{xy}$), and seven $f$ orbitals, ($\phi_{xyz}$, $\phi_{x\alpha}$, $\phi_{y\alpha}$, $\phi_{z\alpha}$, $\phi_{x\beta}$, $\phi_{y\beta}$, $\phi_{z\beta}$), where their functional forms are shown in Appendix~\ref{ap:real_description}.
We consider the atomic-energy level splittings for the $d$ and $f$ orbitals under the CEF and the atomic SOC. 
By assuming that $d$ and $f$ orbitals are located at the on-site position for simplicity, 
the local Hamiltonian $\mathcal{H}_{\rm loc}$ is given by
\begin{align}
\label{eq:locH}
\mathcal{H}_{\rm loc}=\mathcal{H}_{\rm CEF}+ \mathcal{H}_{\rm LS} + \mathcal{H}_{\rm atomic}.  
\end{align}
The first term in Eq.~(\ref{eq:locH}) represents the CEF Hamiltonian. 
We consider the CEF of the polar point group $C_{4{\rm v}}$, which is one of the noncentrosymmetric point groups. 
The CEF Hamiltonian $\mathcal{H}_{\rm CEF}$ consists of the even-parity one $\mathcal{H}_{\rm CEF}^{({\rm even})}$ and odd-parity one $\mathcal{H}_{\rm CEF}^{({\rm odd})}$, which are written as
\begin{align}
\label{eq:CEF}
\mathcal{H}_{\rm CEF} = &\mathcal{H}_{\rm CEF}^{({\rm even})} + \mathcal{H}_{\rm CEF}^{({\rm odd})}, \\
\label{eq:CEFeven}
\mathcal{H}_{\rm CEF}^{({\rm even})} =  &B_{20} O_{20}({\bm r}) + B_{40}  O_{40}({\bm r}) + B_{44} O_{44}^{({\rm c})} ({\bm r})  \notag\\
&  + B_{60}  O_{60}({\bm r}) + B_{64} O_{64}^{({\rm c})} ({\bm r}), \\
\label{eq:CEFodd}
\mathcal{H}_{\rm CEF}^{({\rm odd})} =& B_{10} O_{10}({\bm r}) +  B_{30} O_{30}({\bm r})
+ B_{50} O_{50}({\bm r}) + B_{54} O_{54}^{({\rm c})}({\bm r}),
\end{align}
where $B_{lm}$ ($l=0$-$6$ and $m=0$-$4$) is the CEF parameter and $O_{lm}(\bm{r})=\sqrt{4\pi/(2l+1)}r^l Y^{*}_{lm}(\hat{\bm{r}})$.
$Y_{lm} (\hat{\bm r})$ is the spherical harmonics [$l$ and $m$ are orbital-angular and magnetic quantum numbers ($l \geq 0$, $-l \leq m \leq l$), respectively, and $\hat{\bm r} = {\bm r}/|{\bm r}|$].
In Eqs.~(\ref{eq:CEFeven}) and (\ref{eq:CEFodd}), the tesseral harmonics  are used~\cite{hutchings1964point}:
\begin{align}
O_{lm}^{({\rm c})} ({\bm r}) &\equiv \frac{(-1)^m}{\sqrt{2}} \left[ O_{lm} ({\bm r})+O_{lm}^* ({\bm r}) \right], \\
O_{lm}^{({\rm s})} ({\bm r}) &\equiv \frac{(-1)^m}{\sqrt{2}i} \left[ O_{lm} ({\bm r})-O_{lm}^* ({\bm r}) \right], 
\end{align}
for $l\geq 1$ and $1 \leq |m| \leq l$.
We use $O_{lm} ({\bm r})=(-1)^mO^*_{l-m} ({\bm r})$.
As $O_{lm}(\bm{r})$ has parity $(-1)^l$ with respect to spatial inversion, $\mathcal{H}_{\rm CEF}^{({\rm even})}(\bm{r})=\mathcal{H}_{\rm CEF}^{({\rm even})}(-\bm{r})$ and $\mathcal{H}^{({\rm odd})}_{\rm CEF}(\bm{r})=-\mathcal{H}^{({\rm odd})}_{\rm CEF}(-\bm{r})$ are satisfied.
The even- and odd-parity CEFs in Eqs.~(\ref{eq:CEFeven}) and (\ref{eq:CEFodd}) give rise to qualitatively different splittings;
the former leads to energy splittings between orbitals with the same parity, i.e., $d$-$d$ and $f$-$f$ orbitals, while the latter leads to a mixing between orbitals with the different parity, i.e., $d$-$f$ orbitals~\cite{hanzawa2011crystalline}. 
The second term in Eq.~(\ref{eq:locH}) is the atomic SOC Hamiltonian, which is given by 
\begin{align}
\label{eq:Hls}
\mathcal{H}_{\rm LS} &= \sum_{\zeta=d,f} \frac{\lambda_{\zeta}}{2} {\bm l} \cdot \bm{\sigma},
\end{align}
where $\lambda_{\zeta}$ are the SOC constants for $d$ and $f$ orbitals.
${\bm l}$ and $\bm{\sigma}/2$ are the orbital and spin angular-momentum operators, respectively.
The term in Eq.~(\ref{eq:Hls}) mixes the orbitals with the same parity but different angular momenta. 
The third term in Eq.~(\ref{eq:locH}) represents the atomic energy difference between $d$ and $f$ orbitals where we set an appropriate value below. 

In the situation where the energy scale of the even-parity CEF is larger than those of the odd-parity CEF and the SOC~\cite{comment_CEF}, 
the five $d$ orbitals are split into three single orbitals $\phi_u$ belonging to the irreducible representation ${\rm A_{1}}$, $\phi_v$ to ${\rm B_{1}}$, and $\phi_{xy}$ to ${\rm B_{2}}$ and a doubly-degenerate orbital $(\phi_{yz}, \phi_{zx})$ to ${\rm E}$, while seven $f$ orbitals are into three single orbitals $\phi_{z\alpha}$ to ${\rm A_{1}}$, $\phi_{z\beta}$ to ${\rm B_{1}}$, and $\phi_{xyz}$ to ${\rm B_{2}}$ and two doubly-degenerate orbitals ($\phi_{x\alpha}, \phi_{y\alpha}$) and ($\phi_{x\beta}, \phi_{y\beta}$) to ${\rm E}$.
The schematic level splittings under the even-parity CEF are shown in Fig.~\ref{fig:CEF_split}. 
Note that the even-parity CEF in Eq.~(\ref{eq:CEFeven}) corresponds to that in the centrosymmetric point group $D_{\rm 4h}$. 
Here and hereafter, we focus on two orbitals $\phi_{xy}$ and $\phi_{xyz}$ belonging to the same irreducible representation B$_2$.
The following results are straightforwardly applied to the other orbitals belonging to the different representations, such as A$_1$ and B$_1$.

Next, we consider the effect of the SOC.
The spinful basis $\phi_{xy, \sigma}$ and $\phi_{xyz, \sigma}$ ($\sigma=\uparrow, \downarrow$) 
belong to the double group representation ${\rm E}_{3/2}$ due to ${\rm B}_2 \otimes {\rm E}_{1/2} \to {\rm E}_{3/2}$. 
Considering the mixing with higher energy levels in the first-order perturbation with respect to $\lambda$, ($\phi_{xy, \sigma}, \phi_{xyz, \sigma}$) turns into $(\phi_{d, \sigma},\phi_{f, \sigma})$, which are given by
\begin{align}
\label{eq:base}
{\phi}_{d, \sigma} &=N_d \left[ \phi_{xy, \sigma} \mp 2 i 
 \Lambda^d_{\rm B_1} \phi_{v,\sigma} \pm \Lambda^d_{\rm E} (\phi_{yz, \tilde{\sigma}} \pm i\phi_{zx,\tilde{\sigma}}) \right], \notag \\
{\phi}_{f ,\sigma} &= N_f \left[  \phi_{xyz,\sigma} \mp 2i\Lambda^f_{\rm B_1}\phi_{z\beta, \sigma}
+ i\left(\tilde{\Lambda}_{\rm E}^{(f,1)} \phi_{{\rm E_{3/2}}, \tilde{\sigma}}^{(1)} - \tilde{\Lambda}_{\rm E}^{(f,2)} \phi_{{\rm E_{3/2}, \tilde{\sigma}}}^{(2)} \right)
\right],
\end{align}
where the coefficients are represented by
\begin{align}
\label{eq:lambda}
\Lambda_{\Gamma}^\zeta &= \frac{\lambda_\zeta}{\Delta_{\rm B_2}^{\zeta} - \Delta_{\Gamma}^{\zeta}},\\
\label{eq:lambdaE}
\left(\tilde{\Lambda}_{\rm E}^{(f,1)}, \tilde{\Lambda}_{\rm E}^{(f,2)} \right) &=\left(\frac{2\sqrt{2}\lambda_f}{\Delta_{\rm B_2}^{f} -  \Delta_{\rm E}^{(f,1)}} c_1, 
\frac{2\sqrt{2}\lambda_f}{\Delta_{\rm B_2}^{f} -  \Delta_{\rm E}^{(f,2)}} c_2\right).
 \end{align}
In Eq.~(\ref{eq:base}), $N_d$ and $N_f$ are normalization factors and the subscript $\tilde{\sigma}$ represents the opposite spin to $\sigma$.
$\Delta_{\Gamma}^{\zeta}$ is the atomic energy level split under the even-parity CEF for $\zeta=d, f$ and $\Gamma = {\rm B_1, B_2, E}$.
$\phi_{{\rm E_{3/2}},\sigma}^{(i)}$ with the atomic energy $\Delta_{\rm E}^{(f,i)}$ ($i=1,2$) in Eq.~(\ref{eq:base})  is constructed from linear combinations of $(\phi_{x\alpha,\sigma}, \phi_{y\alpha,\sigma},\phi_{x\beta,\sigma},\phi_{y\beta,\sigma})$.
$c_1$ and $c_2$ in Eq.~(\ref{eq:lambdaE}) are determined by the even-parity CEF parameters.

Finally, let us consider $\mathcal{H}^{\rm (odd)}_{\rm CEF}$ for the basis $(\phi_{d, \sigma},\phi_{f, \sigma})$ 
in Eq.~(\ref{eq:base}). 
As $\mathcal{H}^{\rm (odd)}_{\rm CEF}$ leads to the parity mixing between $d$ and $f$ orbitals, the off-diagonal matrix element between $\phi_{d, \sigma}$ and $\phi_{f, \sigma}$ becomes nonzero, which is evaluated from Eq.~(\ref{eq:CEFodd}) as 
\begin{align}
\label{eq:exp_odd}
\langle \phi_{d,\sigma}|\mathcal{H}^{\rm (odd)}_{\rm CEF} |\phi_{f,\sigma}\rangle
=\frac{33B_{10} - 22B_{30} + 5B_{50} - 3\sqrt{35}B_{54}}{33\sqrt{7}}.
\end{align}
Such an odd-parity hybridization plays an important role in describing the atomic-scale MTD degree of freedom, as discussed in the subsequent sections. 

\subsection{Active multipoles}
\label{subsec:multipole}
We examine active multipoles including the MTD degree of freedom in $(\phi_{d, \sigma},\phi_{f, \sigma})$ in Eq.~(\ref{eq:base}).
In order to describe electronic degrees of freedom in the parity-mixing orbitals, we introduce four types of multipoles: 
electric (E) $\hat{Q}_{lm}$, magnetic (M) $\hat{M}_{lm}$, magnetic toroidal (MT) $\hat{T}_{lm}$, and electric toroidal (ET) $\hat{G}_{lm}$ multipoles~\cite{dubovik1986axial, dubovik1990toroid, hayami2018microscopic}. 
The operator expressions of E, M, MT, and ET multipoles are given by~\cite{kusunose2008description, kuramoto2009multipole, santini2009multipolar, hayami2018microscopic, suzuki2018first}
\begin{align}
\label{eq:Emultipole_Q}
\hat{Q}_{lm} &= -e\sum_j O_{lm} (\bm{r}_j), \\
\label{eq:Mmultipole_M}
\hat{M}_{lm} &= -\mu_{\rm B}\sum_j 
\left(\frac{2 \bm{l}_j}{l+1} + \bm{\sigma}_j\right) \cdot \bm{\nabla} O_{lm}(\bm{r}_j),\\
\label{eq:MTmultipole_T}
\hat{T}_{lm} &=  -\mu_{\rm B}\sum_j 
\left[\frac{\bm{r}_j}{l+1} \times \left( \frac{2\bm{l}_j}{l+2} + \bm{\sigma}_j \right)\right]
\cdot \bm{\nabla} O_{lm}(\bm{r}_j), \\
\label{eq:ETmultipole_G}
\hat{G}_{lm} &= -e\sum_{j}\sum_{\alpha\beta}^{x,y,z}g_{l}^{\alpha\beta}({\bm r}_j)\nabla_{\alpha}\nabla_{\beta} O_{lm}(\bm{r}_{j}),
\end{align}
where $e$ and $\mu_{\rm B}$ represent electron charge and the Bohr magneton, respectively. 
$g_{l}^{\alpha\beta}(\bm{r}_{j})=[2 \bm{l}_j/(l+1) + \bm{\sigma}_j]_{\alpha}[\bm{r}_j \times [ 2\bm{l}_j/(l+2) + \bm{\sigma}_j ]/(l+1)]_{\beta}$.
As $O_{lm}({\bm r}_j)$ has the parity $(-1)^l$
under the spatial inversion, and ${\bm l}_j$ and ${\boldsymbol{\sigma}}_j$ are time-reversal odd, 
the spatial inversion ($\mathcal{P}$) and time-reversal ($\mathcal{T}$) properties of E, M, MT, and ET multipoles are represented by
$(\mathcal{P}, \mathcal{T}) = [(-1)^l,+1]$, $[(-1)^{l+1},-1]$, $[(-1)^{l},-1]$, and $[(-1)^{l+1},+1]$, respectively.
Hereafter, we set $-e$ and $-\mu_{\rm B}$ as $1$ for simplicity.

By calculating the matrix elements of the operators $\langle \phi | \hat{X}_{lm} |\phi^\prime \rangle$ [$\phi, \phi^\prime \in (\phi_{d, \sigma},\phi_{f, \sigma})$ and $\hat{X}=\hat{Q}, \hat{M}, \hat{T}, \hat{G}$] in Eqs.~(\ref{eq:Emultipole_Q})-(\ref{eq:ETmultipole_G}), 
we identify active multipoles in the basis $(\phi_{d, \sigma},\phi_{f, \sigma})$. 
For simplicity, we use the notations
$\hat{X}_0$ for a monopole ($l=0$), $(\hat{X}_x, \hat{X}_y, \hat{X}_z)$
for dipoles ($l=1$), 
($\hat{X}_{u}, \hat{X}_{v}, \hat{X}_{yz}, \hat{X}_{zx}, \hat{X}_{xy}$)
for quadrupoles ($l=2$), 
and ($\hat{X}_{xyz}, \hat{X}_{x\alpha}, \hat{X}_{y\alpha}, \hat{X}_{z\alpha}, \hat{X}_{x\beta}, \hat{X}_{y\beta}, \hat{X}_{z\beta}$) 
for octupoles ($l=3$) instead of $\hat{X}_{lm}$~\cite{hayami2018microscopic, hayami2018classification}. 
See Appendix~\ref{ap:real_description} for each detailed expression.

We describe the active sixteen multipoles spanned by $(\phi_{d, \sigma},\phi_{f, \sigma})$ in Eq.~(\ref{eq:base}) by the product of two Pauli matrices 
$\tau_\mu $ and $\sigma_\nu$ for $\mu, \nu=0, x, y, z$; $\tau_\mu$ represents the orbital degree of freedom between $(\phi_{d, \sigma},\phi_{f, \sigma})$ and 
$\sigma_\nu$ represents the spin degree of freedom ($\tau_0$ and $\sigma_0$ are $2\times 2$ identity matrices in each space). 
Among sixteen multipoles, eight multipoles are even-parity, which are activated in the intra-orbital space: 
E monopole $\hat{Q}_0 = \sigma_0 \tau_0$, E quadrupole $\hat{Q}_u = -\sigma_0 (\tau_0 + \tau_z)$, M dipoles $(\hat{M}_x, \hat{M}_y, \hat{M}_z) = (\sigma_x \tau_0, \sigma_y \tau_0, \sigma_z \tau_0)$, M octupole $\hat{M}_{z\alpha} = -\sigma_z  (\tau_0 + \tau_z)$, and MT quadrupoles 
$(\hat{T}_{yz}, \hat{T}_{zx}) = [-\sigma_x (\tau_0 + \tau_z), \sigma_y (\tau_0 + \tau_z)]$.
On the other hand, remaining eight multipoles in the inter-orbital space are the odd-parity multipoles: E dipoles $(\hat{Q}_x, \hat{Q}_y, \hat{Q}_z) = (-\sigma_y \tau_y, \sigma_x \tau_y, \sigma_0 \tau_x)$, M quadrupole $\hat{M}_u = \sigma_z \tau_x$, MT dipoles $(\hat{T}_x, \hat{T}_y, \hat{T}_z) = (-\sigma_y \tau_x, \sigma_x \tau_x, -\sigma_0 \tau_y)$, and ET quadrupole $\hat{G}_u = -\sigma_z \tau_y$ where we renormalize constant coefficients in each multipole.
These active multipoles are summarized in Table~\ref{table:irrep}.

The results indicate that the MTDs $(\hat{T}_x, \hat{T}_y, \hat{T}_z)$ can be described as the parity-mixing degrees of freedom between $d$ and $f$ orbitals.
Especially, we find two types of the MTDs in the present system. 
The one is the in-plane $x$ and $y$ components of the MTD [$(\hat{T}_x$, $\hat{T}_y) = (-\sigma_y \tau_x, \sigma_x \tau_x)$] depending on both the spin and orbital degrees of freedom. 
The other is the out-of-plane $z$ component of the MTD ($\hat{T}_z =  -\sigma_0 \tau_y$) depending on only the orbital degree of freedom.
Such a difference gives rise to different magnetic responses, as discussed in Sect.~\ref{sec:ME}.

Above multipole classification is also understood from the symmetrical analysis.
By decomposing the product of the irreducible representation ${\rm E_{3/2}}$ for ($\phi_{xy,\sigma}, \phi_{xyz,\sigma}$), sixteen irreducible representations are given by
\begin{align}
&({\rm E_{3/2}} \oplus {\rm E_{3/2}}) \otimes({\rm E_{3/2}} \oplus {\rm E_{3/2}})\notag\\
&=(2{\rm A_1^+} \oplus  2{\rm A_2^-} \oplus 2{\rm E^-} )_{\rm intra} \oplus ({\rm A_1^\pm} \oplus {\rm A_2^\pm} \oplus {\rm E^\pm} )_{\rm inter},
\end{align}
where the superscript $\pm$ represents the parity under the time-reversal operation and $(\dots)_{\rm intra}$ [$(\dots)_{\rm inter}$] represents the multipoles activated in intra-(inter-)orbital degrees of freedom.
Considering the correspondence to the results in Table~\ref{table:irrep},
the intra-orbital even-parity multipoles $\hat{Q}_0$ and $\hat{Q}_u$ belong to ${\rm A}_1^+$, $\hat{M}_z$ and $\hat{M}_z^\alpha$ to ${\rm A}_2^-$ , and $(\hat{M}_x, \hat{M}_y)$ and $(\hat{T}_{yz}, \hat{T}_{zx})$ to ${\rm E}^-$. 
Meanwhile, the inter-orbital odd-parity multipoles are classified into $\hat{Q}_z, \hat{T}_z \in {\rm A}_1^{\pm}$,  $\hat{G}_u, \hat{M}_u \in {\rm A}_2^{\pm}$, and $(\hat{Q}_x, \hat{Q}_y), (\hat{T}_x, \hat{T}_y) \in {\rm E}^{\pm}$, respectively.
The results are also presented in Table~\ref{table:irrep}. 

\renewcommand{\arraystretch}{1.2}
\begin{table}[t]
\caption{
Active multipoles in ($\phi_{d,\sigma}, \phi_{f,\sigma}$) orbitals under the point group $C_{\rm 4v}$.
\label{table:irrep}}
\centering
\begin{tabular}{cll}
\multicolumn{3}{l}{(a) intra-orbital multipole degrees of freedom} \\ \hline \hline
irrep. & multipole & Pauli matrix representation\\
\hline 
${\rm A}_{\rm 1}^+$ &  $\hat{Q}_0$ $\propto$ $1$ & $\sigma_0 \tau_0$  \\
  & $\hat{Q}_u$ $\propto$ $(3z^2-r^2)$ & $-\sigma_0 (\tau_0+\tau_z)$ \\
 \hline
 ${\rm A}_{\rm 2}^-$ & $\hat{M}_z$ $\propto$ $\sigma_z$ &  $\sigma_z \tau_0$ \\
  & $\hat{M}_{z\alpha}$ $\propto$
  $(3z^{2}-r^{2} )\sigma_{z}$& $-\sigma_z (\tau_0+\tau_z)$ \\
 \hline
 ${\rm E}^-$ & $(\hat{M}_x, \hat{M}_y)$ $\propto$ $(\sigma_x, \sigma_y)$ & $(\sigma_x \tau_0, \sigma_y \tau_0)$ \\
 & $(\hat{T}_{yz}, \hat{T}_{zx})$ $\propto$ $(z^2 \sigma_x, -z^2 \sigma_y$) & $ [-\sigma_x (\tau_0+\tau_z), \sigma_y (\tau_0+\tau_z))]$ \\
 \hline
 \multicolumn{3}{c}{} \\
 \multicolumn{3}{l}{(b) inter-orbital 
 multipoles degrees of freedom} \\
\hline
\hline
irrep. & multipole  & Pauli matrix representation
\\
\hline 
${\rm A}_{\rm 1}^+$ & $\hat{Q}_z$ $\propto$ $z$ & $\sigma_0 \tau_x$ \\ 
${\rm A}_{\rm 1}^-$  & $\hat{T}_z$ $\propto$ $xl_y - yl_x$ 
& $-\sigma_0 \tau_y$ \\
 \hline
${\rm A}_{\rm 2}^+$ & $\hat{G}_u$ $\propto$ $(xl_y - yl_x)\sigma_z$ & $-\sigma_z \tau_y$ \\
${\rm A}_{\rm 2}^-$  & $\hat{M}_u$ $\propto$ $z\sigma_z$ & $\sigma_z \tau_x$ \\
 \hline
${\rm E}^+$ & $(\hat{Q}_x, \hat{Q}_y)$ $\propto$ $(x, y)$ & $(-\sigma_y \tau_y, \sigma_x \tau_y)$ \\
${\rm E}^-$ & $(\hat{T}_x, \hat{T}_y)$ $\propto$ ($-z\sigma_y, z\sigma_x$) & $(-\sigma_y \tau_x, \sigma_x \tau_x)$ \\ \hline
\end{tabular}
\end{table}
 \renewcommand{\arraystretch}{1}

\subsection{Effective tight-binding model}
\label{subsec:dfmodel}
\begin{figure}[h]
	\centering
	\includegraphics[width=90mm]{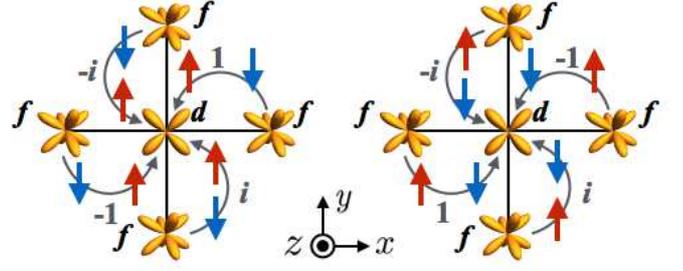} 
	\caption{
	(Color online) Schematic pictures of	the inter-orbital hoppings between $d$ and $f$ orbitals. 
	$\pm1$ and $\pm i$ represent phases of the hybridization.
	\label{fig:hopping}}
	\end{figure}

We construct the tight-binding model on a square lattice where the primitive translation vectors are $(a,0)$ and $(0,a)$ (we set the lattice constant $a=1$).
By adopting the Slater-Koster parameters~\cite{takegahara1980slater} for electron hoppings, the tight-binding Hamiltonian is given by
	\begin{align}
	\label{eq:model_Hamiltonian}
	{\mathcal H}_{df} &= \mathcal{H}_{\rm hopping} + \mathcal{H}_{\rm mixing} + \mathcal{H}_{\rm odd} + \mathcal{H}_{f}, \\
	\label{eq:Hhop}
	\mathcal{H}_{\rm hopping} &= \sum_{\bm k}  \sum_{\zeta}  \sum_{\sigma} \varepsilon_\zeta \left({\bm k} \right) \zeta^\dagger_{{\bm k} \sigma} \zeta_{{\bm k} \sigma}, \\
	\label{eq:Hmix}
	\mathcal{H}_{\rm mixing} &= \sum_{\bm k} \sum_{\sigma, \sigma^\prime}  \left[ \bm{\sigma}^{\sigma \sigma^\prime}  \times \bm{V} \left({\bm k} \right) \right]_z 
	d_{{\bm k} \sigma}^\dagger f_{{\bm k} \sigma^\prime} + {\rm h.c.}, \\
	\label{eq:Hodd}
	\mathcal{H}_{\rm odd} &=  V_{\rm intra} \sum_{\bm k} \sum_{\sigma} d^\dagger_{{\bm k} \sigma} f_{{\bm k} \sigma} + {\rm h.c.},\\
	\label{eq:Hf}
\mathcal{H}_{f} &= \Delta^f \sum_{\bm k} \sum_{\sigma} f^\dagger_{{\bm k}\sigma} f_{{\bm k} \sigma},
	\end{align}
where $\zeta_{{\bm k}\sigma}^\dagger$$(\zeta_{{\bm k}\sigma})$ is a creation (annihilation) operator of an electron for $\zeta=d, f$ with wave number ${\bm k}$ and spin $\sigma=\uparrow, \downarrow$.
The first two terms in Eq.~(\ref{eq:model_Hamiltonian}) describe the off-site components.
$\mathcal{H}_{\rm hopping}$ in Eq.~(\ref{eq:Hhop}) represents the intra-orbital hopping for $d$ and $f$ orbitals where $\varepsilon_\zeta \left({\bm k}\right) = 2t_\zeta ({\rm cos} k_x + {\rm cos} k_y)$. 
$\mathcal{H}_{\rm mixing}$ in Eq.~(\ref{eq:Hmix})
describes the off-site spin-dependent hybridization between $d$ and $f$ orbitals where ${\bm V}\left({\bm k}\right)=2V_{\rm inter} ({\rm sin} k_x, {\rm sin}k_y)$.
Note that 
$V_{\rm inter}$ originates from the atomic SOC, which leads to magnetic anisotropy, and is proportional to $V_{df\pi} \tilde{\Lambda}^{(f,i)}_{\rm E}$ ($i=1,2$) 
($V_{df\pi}$ is the Slater-Koster parameter). 
The bond-dependent phase factors in $\mathcal{H}_{\rm mixing}$ are shown in Fig.~\ref{fig:hopping}.
The third and fourth terms in Eq.~(\ref{eq:model_Hamiltonian}) describe the local part related with Eq.~(\ref{eq:locH}) in Sect.~\ref{subsec:CEF}.
$\mathcal{H}_{\rm odd}$ in Eq.~(\ref{eq:Hodd})
represents the odd-parity CEF Hamiltonian inducing the on-site hybridization between $d$ and $f$ orbitals, where $V_{\rm intra}$ is represented by
 the odd-parity CEF parameter in Eq.~(\ref{eq:exp_odd}).
$\mathcal{H}_{f}$ in Eq.~(\ref{eq:Hf}) 
represents the $f$-orbital energy level measured from the $d$-orbital energy level. 
Hereafter, we fix the parameters as $t_d = 1$ and $t_f=0.1$, and change $V_{\rm inter}$, $V_{\rm intra}$, and $\Delta^f$ in order to examine the effect of the SOC and the odd-parity CEF on the stability of the atomic-scale MTDs and their ME responses.

The tight-binding model in Eq.~(\ref{eq:model_Hamiltonian}) implicitly includes the Rashba-type antisymmetric spin-orbit interaction (ASOI), which is present under the polar-type inversion symmetry breaking~\cite{rashba1960properties}.
In fact, the Rashba-type ASOI, $\sigma_x{\rm sin} k_y-\sigma_y{\rm sin} k_x$, is obtained by performing an appropriate unitary transformation for the basis $(\phi_{d,\sigma}, \phi_{f,\sigma})$ in the presence of both $V_{\rm inter}$ and $V_{\rm intra}$, as shown in Appendix~\ref{ap:ASOI}.
Thus, the model in Eq.~(\ref{eq:model_Hamiltonian}) gives rise to the spin splitting in the band structure. 
Moreover, it is noteworthy that there is no intra-orbital ASOI for $d$ and $f$ orbitals due to the higher order with respect to the odd-parity CEF, as shown in Appendix~\ref{ap:ASOI}.  

For later convenience, we introduce the atomic-scale MTD moments at zero temperature in the model in Eq.~(\ref{eq:model_Hamiltonian}) as
	\begin{align}
	\label{eq:Tx_exp}
T_x 	&\equiv  \braket{\hat{T}_x} =
	-i\braket{d_{{\bm k}\uparrow}^\dagger f_{{\bm k}\downarrow} + f_{{\bm k}\uparrow}^\dagger d_{{\bm k}\downarrow} 
	 - f_{{\bm k}\downarrow}^\dagger d_{{\bm k}\uparrow} - d_{{\bm k}\downarrow}^\dagger f_{{\bm k}\uparrow}} ,\\
	\label{eq:Ty_exp}
T_y	&\equiv \braket{\hat{T}_y} =
	\braket{d_{{\bm k}\uparrow}^\dagger f_{{\bm k}\downarrow} + f_{{\bm k}\uparrow}^\dagger d_{{\bm k}\downarrow}
	+ f_{{\bm k}\downarrow}^\dagger d_{{\bm k}\uparrow} + d_{{\bm k}\downarrow}^\dagger f_{{\bm k}\uparrow}} ,\\
		\label{eq:Tz_exp}
T_z &\equiv \braket{\hat{T}_z}  =
		-i\braket{d_{{\bm k}\uparrow}^\dagger f_{{\bm k}\uparrow} + d_{{\bm k}\downarrow}^\dagger f_{{\bm k}\downarrow}
		- f_{{\bm k}\uparrow}^\dagger d_{{\bm k}\uparrow}- f_{{\bm k}\downarrow}^\dagger d_{{\bm k}\downarrow}}.
	\end{align}
The braket $\braket{}$ is defined as
\begin{align}
\label{eq:def_exp}
\braket{\hat{X}} &\equiv 
\frac{1}{2N_{\bm k}} \sum_{\bm k} \sum_q \braket{q{\bm k}|\hat{X}|q{\bm k}} \theta(\xi_q({\bm k}) - \xi_{\rm F}),
\end{align}
where $N_{\bm k}$ is the number of ${\bm k}$, and 
$\ket{q{\bm k}}$ 
is the eigenvector for the eigenenergy $\xi_{q}({\bm k})$ for the band indices $q$ and ${\bm k}$.
$\xi_{\rm F}$ represents the Fermi energy and $\theta({\xi})$ is the Heaviside step function.
Hereafter, we use the notation $X\equiv \braket{\hat{X}}$ for other multipoles.

\section{Stability}
\label{sec:stability}
We examine when and how the MTDs in Eqs.~(\ref{eq:Tx_exp})-(\ref{eq:Tz_exp}) become nonzero from the energetics point of view. 
In Sect.~\ref{subsec:MF}, we show that the in-plane MTDs $T_x$ and $T_y$ are activated by the spontaneous ferromagnetic state under the intra-orbital Coulomb interaction.
In Sect.~\ref{subsec:Ecomparison}, we discuss the stability of the out-of-plane MTD $T_z$ in the presence of the inter-orbital Coulomb interaction.

\subsection{In-plane spin-dependent MTD}
\label{subsec:MF}
We investigate how to activate the in-plane MTDs $T_x$, $T_y$ in Eqs.~(\ref{eq:Tx_exp}) and (\ref{eq:Ty_exp}) in the model in Eq.~(\ref{eq:model_Hamiltonian}).
We focus on the fact that $T_x$ and $T_y$ belong to the same irreducible representation to $M_y$ and $M_x$, as shown in Table~\ref{table:irrep}, which means that $T_x$ ($T_y$) is indirectly
 induced as a secondary order parameter once the ferromagnetic state with the $M_y$ ($M_x$) component is stabilized in the presence of the odd-parity CEF $V_{\rm intra}$.

In order to examine the stability of the in-plane ferromagnetic ordering, we introduce the Coulomb interaction for the $f$ orbital:
\begin{align}
\label{eq:Coulomb}
\mathcal{H}_{U}
&= U
\sum_{i} f_{i\uparrow}^\dagger f_{i\uparrow} f_{i\downarrow}^\dagger f_{i\downarrow}.
\end{align}
We apply the standard Hartree-Fock approximation as
\begin{align}
\label{eq:HF}
f_{i\uparrow}^\dagger f_{i\uparrow} f_{i\downarrow}^\dagger f_{i\downarrow} \rightarrow 
\left( {Q}_0^f \hat{Q}_0^f - {\bm M}^f \cdot \hat{\bm M}^f  \right) + {\rm const.},
\end{align}
where $\hat{Q}_0^f$ and $\hat{\bm M}^f$ are defined as $\hat{Q}_0^f = \sum_\sigma f_{i\sigma}^\dagger f_{i\sigma}$ and $\hat{\bm M}^f = \sum_{\sigma, \sigma^\prime} f_{i\sigma}^\dagger {\bm \sigma}^f f_{i\sigma^\prime}$, respectively
[$\sigma^f_\mu= (1/2)\sigma_{\mu}(\tau_0-\tau_z)$ for $\mu=x,y,z$]. 

For the mean-field calculations, we consider a single-site unit cell to examine the local magnetic anisotropy, and calculate the mean fields by taking the $\bm{k}$ summation over $200 \times 200$ grid points in the first Brillouin zone.
The mean fields are determined self-consistently within a precision less than $10^{-8}$.
We set $U=10$ and $V_{\rm inter}=0.5$ for the following calculations.

\begin{figure}[htbp]
	\centering
	\includegraphics[width=70mm]{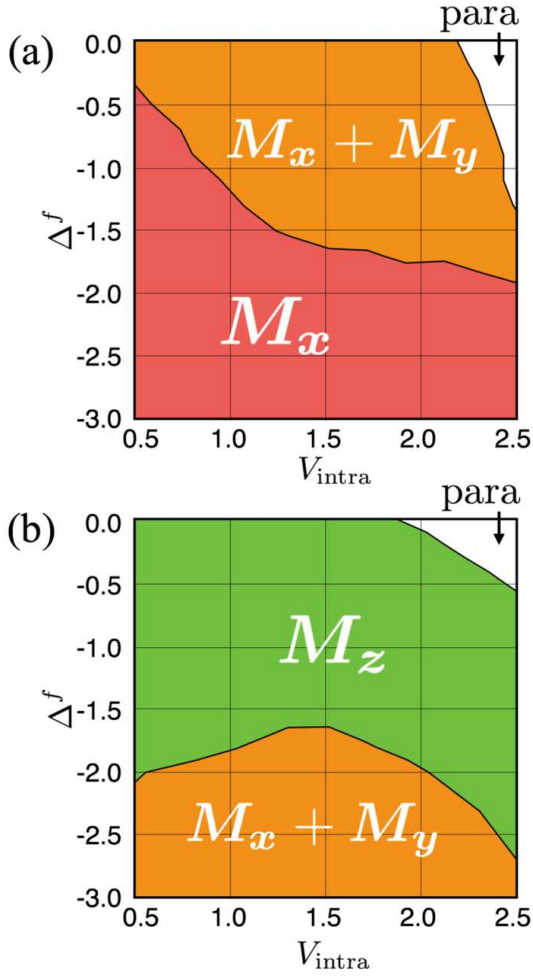}
	\caption{
	(Color online) The preferred magnetic moment direction for the model in Eq.~(\ref{eq:model_Hamiltonian}) 
	in the presence of the intra-orbital Coulomb interaction in Eq.~(\ref{eq:Coulomb}). 
	The phases are obtained by the mean-field calculations at (a) half (1/2) filling and (b) 2/5 filling.
	$M_x$, $M_x+M_y$, $M_z$, and para in the figure represent the magnetic states with the magnetization along the $\langle 100 \rangle$, $\langle 110 \rangle$, and  $\langle 001 \rangle$ direction and the paramagnetic state, respectively.
	\label{fig:MF}}
	\end{figure}

Figure~\ref{fig:MF}(a) shows the zero-temperature phase diagram by changing $V_{\rm intra}$ and $\Delta^f$ at half (1/2) filling. 
There are two in-plane magnetic states with $M_x$ or $M_x+M_y$ in the phase diagram.
The $M_x$ state has nonzero magnetization
along the $\langle 100 \rangle$ direction, while the $M_{x}+M_{y}$ state has nonzero magnetization along the $\langle110\rangle$ direction. 
The $M_x$ state becomes stable in the region for large negative $\Delta^{f}$, while the $M_x+M_y$ state appears in the region for small $\Delta^{f}$.
Both magnetic states are metallic, while the paramagnetic state is insulating with a hybridization gap between $d$ and $f$ orbitals.
The phase transition between two magnetic phases is of first order.
On the other hand, the phase transition between the $M_x+M_y$  and paramagnetic state is of second order.

Both in-plane magnetic states are accompanied with the nonzero MTD moments; 
$-T_y$ ($T_x$) is induced in the $M_x$ ($M_y$) state, whereas $T_x-T_y$ ($-T_x-T_y$) is induced in the $M_x+M_y$ ($M_x - M_y$) state.
The in-plane MTD moments are developed by increasing $V_{\rm intra}$ and decreasing $\Delta^f$, 
while in-plane M dipole moments are suppressed by increasing $V_{\rm intra}$.
This result indicates that the in-plane MTD is affected by the odd-parity CEF $V_{\rm intra}$ as well as the ferromagnetic moment.
The amplitude of $(T_x, T_y)$ is roughly scaled by the products of $(M_y, -M_x)$ and $Q_z$, where $Q_z$ becomes nonzero for nonzero $V_{\rm intra}$. 

The preference of the in-plane magnetic states rather than the out-of-plane magnetic states is presumably understood from the effective inter-orbital Rashba-type ASOI, which consists of the odd-parity CEF $V_{\rm intra}$ and the spin-dependent hopping $V_{\rm inter}$~\cite{hayami2015spontaneous}. 
This tendency to stabilize the in-plane magnetic states is different from the intra-orbital Rashba-type ASOI neglected in the present study, which favors the out-of-plane magnetic anisotropy. 
In fact, we confirmed that the out-of-plane magnetic state is stabilized by introducing the intra-orbital Rashba-type ASOI.
In other words, the inter-orbital ASOI in ``multi-orbital" systems is a key ingredient to stabilize the spin-dependent in-plane MTD state. 

Figure~\ref{fig:MF}(b) shows the phase diagram at a different filling fraction (2/5 filling).
In contrast to the case at half filling, the $M_z$ state with the out-of-plane magnetization is stabilized in the small $\Delta^f$ region.
The $M_z$ state exhibits the M quadrupole $M_u$ because they belong to the same irreducible representation ${\rm A_2^-}$ in Table~\ref{table:irrep}.
In the large negative $\Delta^f$ region, the $M_z$ state is replaced by the $M_x+M_y$ state with the in-plane MTD.
The phase transition between the $M_x+M_y$ and $M_z$ phases is of first order, whereas that between the $M_z$ and paramagnetic state is of second order.
For the case in Fig.~\ref{fig:MF}(b), the in-plane magnetic state is also destabilized by introducing the intra-orbital Rashba-type ASOI.

\subsection{Out-of-plane spin-independent MTD} 
\label{subsec:Ecomparison}
Next, we discuss the stability of the out-of-plane MTD $T_z$ in the model in Eq.~(\ref{eq:model_Hamiltonian}).
In contrast to ($T_x$, $T_y$), any magnetic orders do not induce $T_z$, as they are not coupled with each other. 
In order to realize the $T_z$ ordering, we introduce the inter-orbital Coulomb interaction between $d$ and $f$ orbitals, which is represented as
\begin{align}
\label{eq:H_interU}
\mathcal{H}_{U'} = 
U^{\prime} \sum_i \sum_{\sigma, \sigma^\prime} f_{i\sigma}^\dagger f_{i\sigma} d_{i\sigma^\prime}^\dagger d_{i\sigma^\prime}.
\end{align}
The mean-field decoupling for the Fock terms leads to an effective interaction to induce $T_z$, which is represented by
\begin{align}
\label{eq:equal} 
&\sum_{\sigma, \sigma^\prime} \left[f_{i\sigma}^\dagger f_{i\sigma} d_{i\sigma^\prime}^\dagger d_{i\sigma^\prime}\right]_{\rm Fock} 
\nonumber \\
&\to 
-\frac{1}{2} \left( \bm{T} \cdot \hat{\bm{T}} + \bm{Q} \cdot \hat{\bm{Q}} + G_u \hat{G}_u + M_u \hat{M}_u \right) + {\rm const.}, 
 \end{align}
where each multipole operator is constructed by using $\tau_\mu \sigma_\nu$ ($\mu,\nu=0,x,y,z$) in Table~\ref{table:irrep}.
When the Fock term in Eq.~(\ref{eq:equal}) is a dominant interaction, one of eight multipoles $(T_x, T_y, T_z, Q_x, Q_y, Q_z, G_u, M_u)$ in the inter-orbital space is activated. 
Note that energies for eight multipole states are degenerate under the Fork term.
 
\begin{figure}[htbp]
	\begin{center}
	\includegraphics[width=90mm]{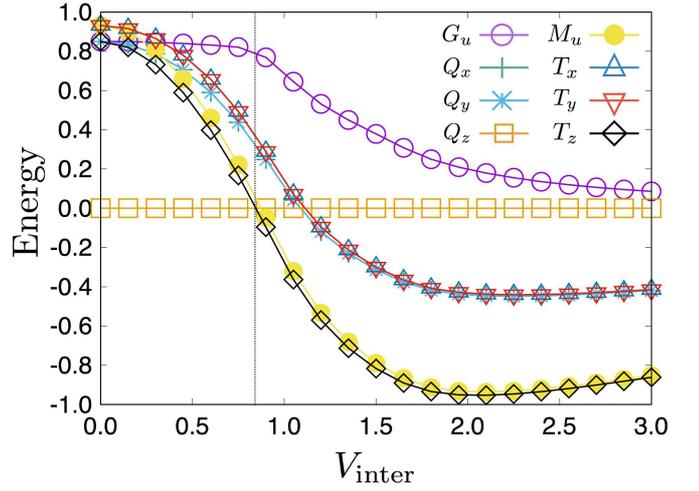} 
	\end{center}
	\caption{
	(Color online) The energy at $T=0$ for each ordered state measured from that for the $Q_z$ state as a function of $V_{\rm inter}$.
	The data are at $V_{\rm intra}=0.5$, $U^{\prime}=10$, and half filling.
	The vertical dashed line at $V_{\rm intra} \sim 0.85$ shows the boundary between the $Q_z$ and $T_z$ states.
	  \label{fig:energy}}
	\end{figure}

The energy degeneracy for eight multipoles is lifted by considering $V_{\rm intra}$ and $V_{\rm inter}$.
Figure~\ref{fig:energy} shows $V_{\rm inter}$ dependences of the energies per site for eight states measured from that for the $Q_z$ state at $U^\prime =10$, $V_{\rm intra}=0.5$, and half filling. 
We assume the saturated mean-field values for each multipole state.
At $V_{\rm inter}=0$, the $Q_z$ state has the lowest energy among the eight states.
While increasing $V_{\rm inter}$, the energy difference between the $Q_z$ state and the $T_z$ state is smaller, and the $T_z$ state becomes the lowest energy state for $V_{\rm inter} \gtrsim 0.85$. 
While further increasing $V_{\rm inter}$, the energy for the $M_u$ state is close to that for the $T_z$ state, and they are almost degenerate for $V_{\rm inter} \gtrsim 3.0$.
The energy difference between the $T_z$ and $M_u$ states is determined by $V_{\rm intra}$. 

The result indicates that the out-of-plane $T_z$ state is realized in the system for the intermediate region of the inter-site hybridization $V_{\rm inter}$ and nonzero $V_{\rm intra}$, although the self-consistent calculations are required to settle this point.
The excitonic insulators, which have been often characterized by the off-diagonal orbital orderings, might be potential candidates to exhibit the $T_z$ state~\cite{kaneko2016electric}.

\section{Magneto-Electric Effect}
\label{sec:ME}
Let us investigate the ME effect under the MTD orderings.
Before discussing the results, we briefly review the ME effect from the symmetrical point of view~\cite{hayami2018classification, watanabe2018group}.
The ME effect where the uniform magnetization $\bm{M}$ is induced by an external electric 
field $\bm{E}$ is expressed as
\begin{align}
\label{eq:ME_tensor_eq}
{\bm M} = \hat{\alpha} {\bm E},
\end{align}
where $\hat{\alpha}$ is the ME tensor.
$\hat{\alpha}$ is calculated by the linear response theory~\cite{yanase2013magneto}; 
\begin{align}
\label{eq:ME_1}
\alpha_{\mu \nu} = \frac{eg \mu_{\rm B}\hbar}{2Vi} \sum_{\bm k} \sum_{p,q} \frac{f[
\xi_p({\bm k})] - f[\xi_q({\bm k})]}{\xi_p({\bm k}) - \xi_q({\bm k})}
 \frac{\sigma_{\mu,{\bm k}}^{pq}\varv_{\nu,{\bm k}}^{qp}}{\xi_p({\bm k}) - \xi_q({\bm k})
+ i\hbar \delta}, 
\end{align}
where $V$ is the system volume, $\hbar=h/2\pi$ is the Plank constant, and $\delta$ is the broadening factor.  
$f[\xi({\bm k)}]$
 represents the Fermi distribution function and $\xi({\bm k})$
 is the eigenenergy. 
As $\sigma^{pq}_{\mu,{\bm k}}=\braket{p{\bm k}|\sigma_\mu|q{\bm k}}$ and $\varv^{pq}_{\nu, {\bm k}} =\braket{p{\bm k}|\varv_{\nu,\bm k}|q{\bm k}}$ ($\mu, \nu=x, y, z$)
are matrix elements of spin moment and velocity $\varv_{\mu, {\bm k}}=\partial {\mathcal H}/ ( \hbar \partial k_{\mu} )$ at ${\bm k}$, $\alpha_{\mu\nu}$ represents the correlation of the magnetic moment and electric current~\cite{yanase2013magneto}.
We take $eg\mu_{\rm B}\hbar/2=1$.

The ME tensor $\alpha_{\mu\nu}$ can be divided into two parts: 
the dissipative part (current-driven) $\alpha_{\mu \nu}^{({\rm J})}$ and  
the non-dissipative part (electric-field driven) $\alpha_{\mu \nu}^{({\rm E})}$. The expressions are shown as
\begin{align}
\label{eq:ME_J}
\alpha_{\mu \nu}^{\rm (J)} &= -\frac{\hbar \delta}{V} \sum_{\bm k} \sum_{p,q}^{=} \frac{f[
\xi_p({\bm k})] - f[\xi_q({\bm k})]}{\xi_p({\bm k}) - \xi_q({\bm k})} \Pi_{\mu \nu}^{pq} ( {\bm k}), \\
\label{eq:ME_E}
\alpha_{\mu \nu}^{\rm (E)} &= \frac{1}{Vi} \sum_{\bm k} \sum_{p,q}^{\neq} 
\{f[\xi_p({\bm k})] - f[\xi_q({\bm k})] \}
\  \Pi_{\mu \nu}^{pq} ( {\bm k}),
\end{align}
where 
$\Pi_{\mu \nu}^{pq} ( {\bm k}) = \sigma_{\mu,{\bm k}}^{pq}\varv_{\nu,{\bm k}}^{qp}/ 
\{ [\xi_p({\bm k}) - \xi_q({\bm k})]^2 + (\hbar \delta)^2\}$. 
The dissipative part $\alpha_{\mu \nu}^{({\rm J})}$ represents the intra-band contribution, whereas the non-dissipative part $\alpha_{\mu \nu}^{({\rm E})}$ represents the inter-band contribution.
We set $\hbar=1$.

Each component of the ME tensors, $\alpha_{\mu \nu}^{({\rm J})}$ and $\alpha_{\mu \nu}^{({\rm E})}$, is related with the odd-parity multipoles, which is represented by~\cite{hayami2018classification}
\begin{align}
\label{eq:alphaJ}
\hat{\alpha}^{({\rm J})} 
&= 
\begin{pmatrix}
\alpha_{xx}^{({\rm J})}  & \alpha_{xy}^{({\rm J})}  & 0 \\
\alpha_{yx}^{({\rm J})}  & \alpha_{yy}^{({\rm J})}  & 0 \\
\alpha_{zx}^{({\rm J})}  & \alpha_{zy}^{({\rm J})}  & 0
\end{pmatrix}
\sim
\begin{pmatrix}
G_{u} & Q_{z} & 0 \\
-Q_{z} & G_{u} & 0 \\
 Q_{y} & - Q_{x} & 0
\end{pmatrix}, \\
\label{eq:alphaE}
\hat{\alpha}^{({\rm E})} 
&= 
\begin{pmatrix}
\alpha_{xx}^{({\rm E})}  & \alpha_{xy}^{({\rm E})}  & 0 \\
\alpha_{yx}^{({\rm E})}  & \alpha_{yy}^{({\rm E})}  & 0 \\
\alpha_{zx}^{({\rm E})}  & \alpha_{zy}^{({\rm E})}  & 0
\end{pmatrix}
\sim
\begin{pmatrix}
M_{u} & T_{z} & 0 \\
-T_{z} & M_{u} & 0 \\
T_{y} &  - T_{x} & 0
\end{pmatrix},
\end{align}
where $\alpha_{\mu z}^{({\rm J})} = \alpha_{\mu z}^{({\rm E})} =0$ in the two-dimensional system.
The paramagnetic state in the model in Eq.~(\ref{eq:model_Hamiltonian}) exhibits nonzero responses $\alpha_{xy}^{({\rm J})}=-\alpha_{yx}^{({\rm J})}$ due to the nonzero odd-parity CEF, resulting in the E dipole $Q_z$. 
The other components become nonzero when the corresponding multipoles are activated.

It is note that the ME tensors $\alpha_{\mu \nu}^{({\rm J})}$ and $\alpha_{\mu \nu}^{({\rm E})}$ are
also active for different multipole ordered states according to the crystallographic point-group symmetry.
In the present model under the point group $C_{\rm 4v}$, $\alpha_{zy}^{({\rm E})}$ and $\alpha_{zx}^{({\rm E})}$ become nonzero when the M dipole ($M_y$, $M_x$) and/or the MT quadrupole ($T_{zx}$, $T_{yz}$) are activated, since they belong to the same irreducible representation E$^-$ as the MT dipole ($T_x$, $T_y$) shown in Table~\ref{table:irrep}.
In a similar way, $\alpha_{xx}^{({\rm E})} =\alpha_{yy}^{({\rm E})}$ in Eq.~(\ref{eq:alphaE}) also becomes nonzero in the presence of $M_z$ in Table~\ref{table:irrep}.

In the following, we focus on the ME tensor $\alpha_{\mu\nu}^{\rm (E)}$ induced by the two types of MTDs.
The results for the in-plane MTDs $T_x, T_y$ are shown in Sect.~\ref{subsec:METy}, and those for the out-of-plane MTD $T_z$ are shown in Sect.~\ref{subsec:METz}.

\subsection{In-plane spin-dependent MTD}
\label{subsec:METy}
We investigate the ME effect induced by the in-plane spin-dependent MTDs, $T_x$ and $T_y$.
As discussed in Sect.~\ref{subsec:MF}, 
since ($T_x, T_y$) is induced under the ($M_y, M_x$) state, we introduce an in-plane magnetic field to the model in Eq.~(\ref{eq:model_Hamiltonian}). 
We consider the magnetic field along the $x$ direction, which is given by
\begin{align}
\label{eq:Hamiltonian_Ty}
\mathcal{H} = 
- H_x \sum_{\bm k} \hat{M}_x,
\end{align}
where $\hat{M}_x = \sum_{\zeta \zeta^\prime} \sum_{\sigma \sigma^\prime} \zeta_{{\bm k}\sigma}^\dagger \sigma_x^{\sigma \sigma^\prime}\tau_0^{\zeta \zeta^\prime} \zeta^\prime_{{\bm k}\sigma^\prime}$ for $\zeta=d, f$ and $\sigma=\uparrow$, $\downarrow$. 
The $g$ factors for $d$ and $f$ orbitals are taken as the same for simplicity.

\begin{figure}[t]
	\centering
	\includegraphics[width=85mm]{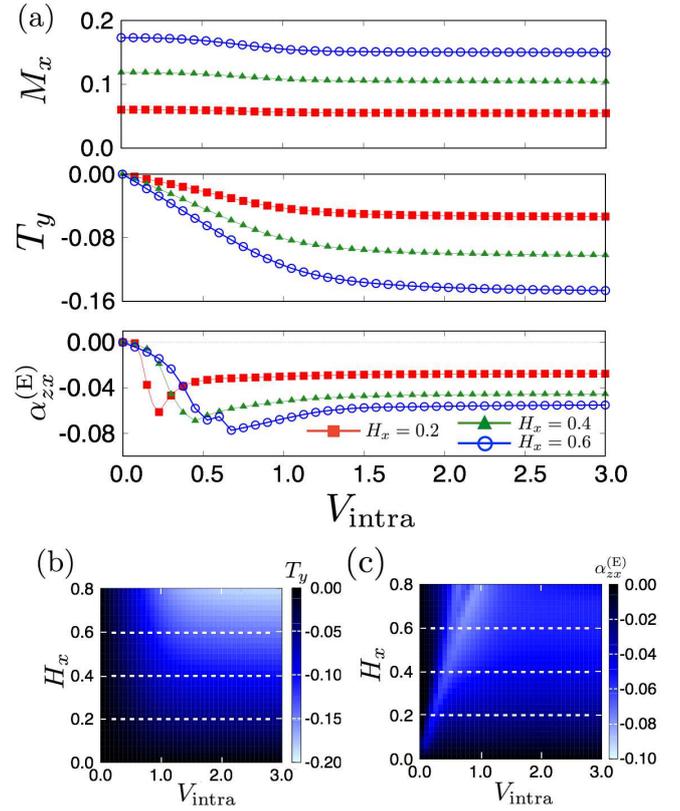} 
	\caption{ 
	(Color online) (a) $V_{\rm intra}$ dependences of  $M_x$, $T_y$, and $\alpha_{zx}^{({\rm E})}$ for $H_x=0.2, 0.4, 0.6$.
	(b), (c) Counter plots of (b) $T_y$ and (c) $\alpha_{zx}^{({\rm E})}$ in the plane of $H_x$ and $V_{\rm intra}$.
	The dashed lines on (b) and (c) correspond to the data presented in (a).
	The other parameters are taken at $\Delta^f=-1.0$, $V_{\rm inter}=0.5$, $T=0.1$, $\delta = 0.01$, and 1/5 filling. 
	\label{fig:METy}}
	\end{figure}

Figure~\ref{fig:METy}(a) shows the M dipole moment $M_x$, the MT dipole moment $T_y$, and the ME tensor $\alpha_{zx}^{({\rm E})}$ for $H_x = 0.2, 0.4, 0.6$ as a function of $V_{\rm intra}$.
The parameters are taken at $\Delta^f=-1$, $V_{\rm inter}=0.5$, $T=0.1$, $\delta = 0.01$, and 1/5 filling.
As shown in the top panel of Fig.~\ref{fig:METy}(a), $M_x$ is larger with an increase of the magnetic field $H_x$, whereas it is almost independent of $V_{\rm intra}$.
On the other hand, $T_y$ becomes larger while increasing either $H_x$ or $V_{\rm intra}$, as shown in the middle panel of Fig.~\ref{fig:METy}(a).
Note that $T_y=0$ for $V_{\rm intra}=0$.
The ME tensor $\alpha_{zx}^{({\rm E})}$, as shown in the bottom panel of Fig.~\ref{fig:METy}(a), exhibits a more complicated behavior.
For $H_x=0.4$, 
it becomes larger with an increase of $V_{\rm intra}$ for small $V_{\rm intra}$.
While further increasing $V_{\rm intra}$, it shows a kink at $V_{\rm intra} \sim 0.45$, and 
takes a constant value for $V_{\rm intra} \gtrsim 1.8$. 
Such a behavior of $\alpha_{zx}^{({\rm E})}$ is qualitatively common to other magnetic fields, $H_x=0.2$ 
and $0.6$.

\begin{figure}[htbp]
	\centering
	\includegraphics[width=90mm]{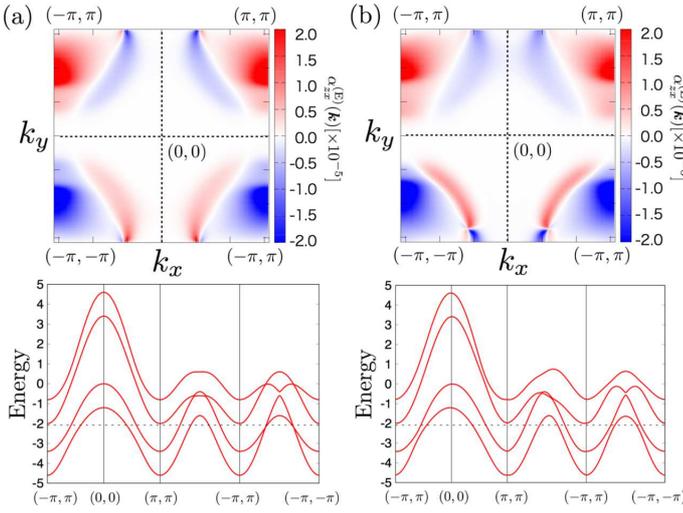} 
	\caption{
	 (Color online) (Upper panel) The ME tensor $\alpha_{zx}^{({\rm E})}(\bm{k})$ decomposed at each wave number in the first Brillouin zone for (a) $V_{\rm intra}=0$ and (b) $V_{\rm intra}=0.1$. 
	The other parameters are $H_x=0.4$, 
	 $\Delta^f=-1.0$, and $V_{\rm inter}=0.5$. 
	(Lower panel) The corresponding energy bands. 
	The dashed lines show the Fermi levels.
	\label{fig:band}}
	\end{figure}

In order to examine $\alpha_{zx}^{({\rm E})}$, we discuss the relationship with the band structure.
We decompose the ME tensor $\alpha_{zx}^{({\rm E})} $ into each wave number defined as $\alpha_{zx}^{({\rm E})}({\bm k}) = \sum_{p,q}^{\neq} {\{f[\xi_p({\bm k})] - f[\xi_q({\bm k})] \}} \Pi_{zx}^{pq} ( {\bm k})/(Vi)$. 
Figures~\ref{fig:band}(a) and (b) represent $\alpha_{zx}^{({\rm E})}({\bm k})$
at $V_{\rm intra}=0$ [Fig.~\ref{fig:band}(a)] and at $V_{\rm intra}=0.1$ [Fig.~\ref{fig:band}(b)] for fixed $H_x=0.4$.
The corresponding band dispersions are also shown in the lower panels of Fig.~\ref{fig:band}. 
At $V_{\rm intra}=0$, the energy band shows a symmetric structure with respect to the wave number $(k_x, k_y) \leftrightarrow (-k_x, -k_y)$ because $Q_z$ resulting from the odd-parity CEF is not activated.
Meanwhile, $\alpha_{zx}^{({\rm E})}({\bm k})$ shows a symmetric structure with respect to $k_x$, $\alpha_{zx}^{({\rm E})}(k_x, k_y)=\alpha_{zx}^{({\rm E})}(-k_x, k_y)$, while it shows an antisymmetric structure with respect to $k_y$, $\alpha_{zx}^{({\rm E})}(k_x, k_y)=-\alpha_{zx}^{({\rm E})}(k_x, -k_y)$, as shown in the upper panel of Fig.~\ref{fig:band}(a).
The asymmetric structure with respect to $k_y$ is due to the breaking of the mirror symmetry in the $xz$ plane under the magnetic field $H_x$. 
Nevertheless, no MTDs are induced at $V_{\rm intra}=0$, i.e., $\alpha_{zx}^{({\rm E})}= \sum_{\bm{k}}\alpha_{zx}^{({\rm E})}({\bm k}) = 0$, since 
$\alpha_{zx}^{({\rm E})}(k_x, k_y)=-\alpha_{zx}^{({\rm E})}(-k_x, -k_y)$. 
The result is consistent with the absence of $T_y$ at $V_{\rm intra}=0$.
For an infinitesimal $V_{\rm intra}$, the structure $\alpha_{zx}^{({\rm E})}({\bm k})$ shows an asymmetric modulation along the $k_y$ direction, as shown in Fig.~\ref{fig:band}(b)~\cite{hayami2014toroidal}.
This result indicates the emergence of $T_y$, which gives rise to nonzero $\alpha_{zx}^{({\rm E})}$. 

Figures~\ref{fig:METy}(b) and (c) represent the contour plots of $T_y$ and $\alpha_{zx}^{({\rm E})}$ in the $V_{\rm intra}$-$H_x$ plane, respectively.
$T_y$ shows a large value with an increase of either $V_{\rm intra}$ or $H_x$, which is a similar tendency in Fig.~\ref{fig:METy}(a). 
On the other hand, the ME tensor $\alpha_{zx}^{({\rm E})}$ becomes larger with an increase of $V_{\rm intra}$ for small $V_{\rm intra}$, while it takes constant value for large $V_{\rm intra}$.
The position of the kink is roughly scaled as $V_{\rm intra} \sim H_x$, which indicates that a large ME response is expected for $V_{\rm intra} \sim H_x$. 
A further analysis by taking into account the electron correlation beyond the mean-field level might also bring a large ME tensor~\cite{peter2018strong}.

\subsection{Out-of-plane spin-independent MTD}
\label{subsec:METz}
We investigate the ME effect under the out-of-plane MTD $T_z$ ordering, which is expected to show nonzero $\alpha_{xy}^{\rm (E)}=-\alpha_{yx}^{\rm (E)}$.
We introduce the mean filed to induce $T_z$,
which is given by
\begin{align}
\label{eq:Hamiltonian_Tz}
\mathcal{H} = 
- H_z^{({\rm T})} 
\sum_{\bm k} \hat{T}_z,
\end{align}
where $\hat{T}_z = \sum_{\zeta \zeta^\prime} \sum_{\sigma\sigma^\prime} \zeta_{{\bm k}\sigma}^\dagger \sigma_0^{\sigma\sigma^\prime} \tau_y^{\zeta\zeta^\prime} \zeta^\prime_{{\bm k}\sigma^\prime}$ for $\zeta,\zeta^\prime=d, f$ and $\sigma, \sigma^\prime = \uparrow, \downarrow$. 
$H_z^{({\rm T})}$ in Eq.~(\ref{eq:Hamiltonian_Tz}) corresponds to the mean field $U^\prime T_z/2 $ in Sect.~\ref{subsec:Ecomparison}.

Contrary to the symmetrical argument in Eq.~(\ref{eq:alphaE}),
it is found that the ME tensor $\alpha_{xy}^{({\rm E})}$ becomes zero even under the $T_z$ ordering by calculating it.
In order to investigate whether a nonzero $\alpha_{xy}^{({\rm E})}$ can be obtained or not, we examine the behavior of $\alpha_{xy}^{({\rm E})}$ in details. 
For that purpose, we decompose the ME tensor $\alpha_{xy}^{({\rm E})}$ in Eq.~(\ref{eq:ME_E}) into that for $d$($f$)-orbital components, 
which are defined as 
\begin{align}
\label{eq:ME_orbit}
\alpha_{xy}^{\rm (E, \zeta)} &= \frac{1}{Vi} \sum_{\bm k} \sum_{p,q}^{\neq} 
\{f[\xi_p({\bm k})] - f[\xi_q({\bm k})] \}
\  \Pi_{xy}^{(\zeta) pq} ( {\bm k}),
\end{align}
where the total spin moment $\sigma_{\mu,\bm{k}}^{pq}$ in Eq.~(\ref{eq:ME_E}) is replaced with that for the $d$-orbital spin moment $\sigma^{(d)}_x = \sigma_x (\tau_0 + \tau_z)/2$ for $\zeta=d$ and the $f$-orbital spin moment $\sigma^{(f)}_x = \sigma_x (\tau_0 - \tau_z)/2$ for $\zeta=f$.
The nonzero $\alpha_{xy}^{({\rm E},d)}$ [$\alpha_{xy}^{({\rm E},f)}$] in Eq.~(\ref{eq:ME_orbit}) means that the $x$ component of the magnetization for the $d$ ($f$) orbital is induced for an applied electric field along the $y$ direction.
As the relation $\alpha_{xy}^{({\rm E},d)} =-\alpha_{xy}^{({\rm E},f)}$ holds from $\alpha_{xy}^{({\rm E})} =\alpha_{xy}^{({\rm E},d)}+\alpha_{xy}^{({\rm E},f)} =0$ in the present model, we discuss $\alpha_{xy}^{({\rm E},d)}$ in the following.

\begin{figure}[t!]
	\centering
	\includegraphics[width=85mm]{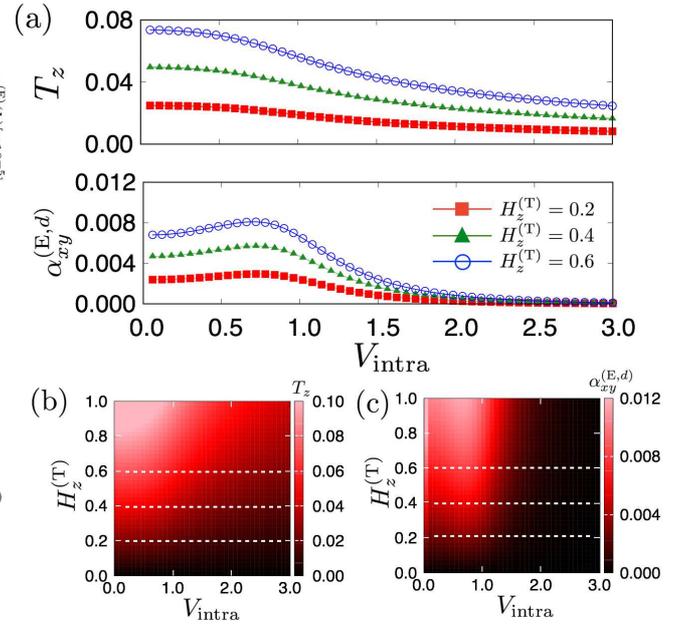} 
	\caption{
	(Color online) (a) $V_{\rm intra}$ dependences of $T_z$ and $\alpha_{xy}^{({\rm E}, d)}$ for $H_z^{\rm (T)}=0.2, 0.4, 0.6$.
	(b), (c) Counter plots of (b) $T_z$ and (c) $\alpha_{xy}^{({\rm E}, d)}$ in the plane of $H_z^{\rm (T)}$ and $V_{\rm intra}$.
	The dashed lines on (b) and (c) correspond to the data presented in (a). The other parameters are
	$\Delta^f=-1.0$, $V_{\rm inter}=0.5$, $T=0.1$, $\delta = 0.01$ and 1/5 filling. 
	\label{fig:METz1}}
	\end{figure}

Figure~\ref{fig:METz1}(a) shows $V_{\rm intra}$ dependences of $T_z$ and the ME tensor $\alpha_{xy}^{({\rm E},d)}$ for $H_{z}^{({\rm T})} = 0.2, 0.4, 0.6$.
$T_z$ shows a similar dependence on $V_{\rm intra}$ for each $H_{z}^{({\rm T})}$.
$T_z$ becomes larger with an increase of $H_z^{({\rm T})}$, whereas it is suppressed while $V_{\rm intra}$ increases.
Meanwhile, the behavior of $\alpha_{xy}^{({\rm E},d)}$ is similar to that of $T_z$ except for small $V_{\rm intra}$.
For small $V_{\rm intra}$, $\alpha_{xy}^{({\rm E},d)}$ becomes slightly larger while increasing $V_{\rm intra}$, and shows a broad peak around $V_{\rm intra} \sim 0.8$.
While further increasing $V_{\rm intra}$, 
$\alpha_{xy}^{({\rm E},d)}$ decreases gradually and approaches to zero. 

In contrast to the ME effect under the $T_y$ ordering in Sect.~\ref{subsec:METy}, both $T_z$ and $\alpha_{xy}^{({\rm E},d)}$ become nonzero at $V_{\rm intra}=0$.
Moreover, the band structure under the $T_z$ ordering is symmetric with respect to $(k_x, k_y)\leftrightarrow (-k_x, -k_y)$ because the asymmetric structure is expected to 
occur along the $k_z$ direction in a three-dimensional system.

Figures~\ref{fig:METz1}(b) and (c) show the contour plots of $T_z$ and $\alpha_{xy}^{({\rm E},d)}$ in the $V_{\rm intra}$-$H_z^{({\rm T})}$ plane. 
$T_z$ shows a large value in the region for small $V_{\rm intra}$ and large $H_z^{({\rm T})}$, which is similar to the result in Fig.~\ref{fig:METz1}(a). 
On the other hand, the ME tensor $\alpha_{xy}^{({\rm E},d)}$ shows a maximum value at $V_{\rm intra}\sim 0.8$ , which is almost independent of $H_{z}^{({\rm T})}$, and decreases while increasing $V_{\rm intra}$ or decreasing $H_z^{({\rm T})}$.

\begin{figure}[t!]
	\centering
	\includegraphics[width=80mm]{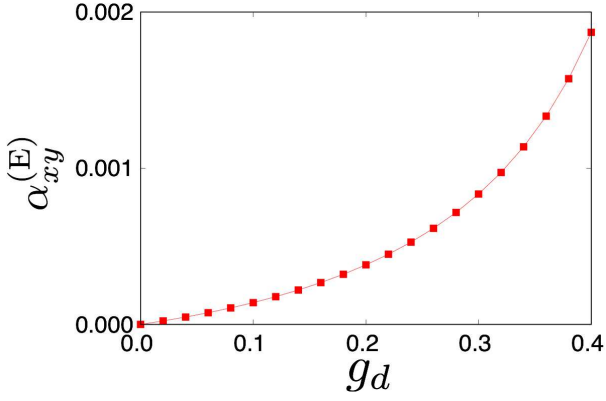} 
	\caption{
	(Color online) The intra-orbital ASOI $g_d(=2 g_f)$ dependence of $\alpha_{xy}^{\rm (E)}$ for $H_z^{\rm (T)}=0.2$, and 
	$V_{\rm intra}=1$. 
	The other parameters correspond to those in Fig.~\ref{fig:METz1}.  
	\label{fig:intra_ASOI}}
	\end{figure}

Meanwhile, the nonzero ME tensor $\alpha_{xy}^{({\rm E})}$ is obtained by taking into account the intra-orbital ASOI in addition to the Hamiltonian in Eq.~(\ref{eq:model_Hamiltonian}), which is neglected as a higher-order contribution compared to the inter-orbital ASOI.
The intra-orbital ASOI Hamiltonian is given by
\begin{align}
\label{eq:intra_ASOI1}
\mathcal{H}_{\rm ASOI}^{\rm intra} =
 \sum_{\bm k} \sum_{\zeta}
\sum_{\sigma,\sigma^\prime}
g_\zeta ({\rm sin}k_x \sigma_y^{\sigma \sigma^\prime} - {\rm sin}k_y \sigma_x^{\sigma\sigma^\prime})
\ \zeta_{{\bm k}\sigma}^\dagger \zeta_{{\bm k}\sigma^\prime},
\end{align}
where $g_\zeta$ is the magnitude of the intra-orbital ASOI for the orbital $\zeta$ ($\zeta=d, f$). 
The derivation of the intra-orbital ASOI in Eq.~(\ref{eq:intra_ASOI1}) is shown in Appendix~\ref{ap:ASOI}.
We show the net component of the ME tensor $\alpha_{xy}^{({\rm E})}$ as a function of the $d$-orbital intra-orbital ASOI $g_d (=2g_f)$ in Fig.~\ref{fig:intra_ASOI}.
The result indicates that $\alpha_{xy}^{({\rm E})}$ becomes nonzero for an infinitesimal $g_d$ and increases with an increase of $g_d$. 
The behaviors of nonzero $\alpha_{xy}^{({\rm E})}$ while changing $H_z^{(\rm T)}$ and $V_{\rm intra}$ are similar to those of $\alpha_{xy}^{({\rm E},d)}$; $\alpha_{xy}^{({\rm E})}$ is enhanced by $H_z^{(\rm T)}$ and suppressed by $V_{\rm intra}$.

\section{Summary}
\label{sec:summary}
We have investigated the atomic-scale MTD originating from the parity-mixing orbital degree of freedom in the microscopic model.
We derived a minimal two-orbital model to examine the nature of the MTD by taking into account the odd-parity CEF and the atomic SOC.
We clarified that two types of the MTDs emerge as an atomic object.
The one is the in-plane MTDs depending on both the spin and orbital degrees of freedom, and the other is the out-of-plane MTD depending on only the orbital degree of freedom.
We found that the in-plane MTD is activated by the in-plane magnetic moments or external magnetic field, while the out-of-plane MTD is expected to occur when the inter-orbital Coulomb interaction becomes large.
Moreover, we examined a behavior of the ME effect for each MTD. 
We have shown that the large (small) odd-parity hybridization is favorable for the large ME effect in the in-(out-of-)plane MTD.

The atomic-scale MTDs in the present study can be realized in other noncentrosymmetric crystals.
For example, the atomic-scale MTD can be induced by appling an external magnetic field to materials with the polar/chiral crystal structure, such as noncentrosymmetric superconductor ${\rm CePt}_3{\rm Si}$~\cite{bauer2004heavy} and ${\rm Ce}T{\rm Si}_3$($T={\rm Rh}, {\rm Ir}$)~\cite{kimura2005pressure, sugitani2006pressure}. 
Furthermore, such a MTD degree of freedom is emergent even in centrosymmetric crystals  
when there is no local inversion symmetry at lattice sites, such as Ce$T$AsO ($T$=Fe, Co, Ni, Mn)~\cite{zhao2008structural, sarkar2010interplay, luo2011ceniaso, zhang2015spin}. 
Our microscopic analysis will give an insight for exploring the atomic-scale MTD in these materials. 

\section*{Acknowledgments}
We thank H. Kusunose, Y. Yanagi, H. Amitsuka, and T. Yanagisawa for fruitful discussions. This research was supported by JSPJ KAKENHI Grants Numbers JP18H04296 (J-Physics) and JP18K13488.

\appendix
\section{Definitions of Atomic $d$- and $f$-Orbital Wave Functions and Atomic Multipoles}
\label{ap:real_description}
In this Appendix, we present the definitions of atomic $d$- and $f$-orbital wave functions.
The five $d$-orbital wave functions ($\phi_u, \phi_v, \phi_{yz}, \phi_{zx}, \phi_{xy}$) are represented by
\begin{align}
\label{eq:uv}
&\phi_u =\frac{1}{2}\sqrt{\frac{5}{4\pi}} \frac{3z^2-r^2}{r^2}, \phi_v = \frac{1}{2} \sqrt{\frac{15}{4\pi}} \frac{x^2-y^2}{r^2},\\
\label{eq:yz}
&(\phi_{yz}, \phi_{zx}, \phi_{xy}) = \sqrt{\frac{15}{4\pi}} \left(\frac{yz}{r^2}, \frac{zx}{r^2}, \frac{xy}{r^2} \right),
\end{align}
and the seven $f$-orbital wave functions ($\phi_{xyz}, \phi_{x\alpha}, \phi_{y\alpha}, \phi_{z\alpha}, \phi_{x\beta}, \phi_{y\beta}, \phi_{z\beta}$) are represented as
\begin{align}
\label{eq:xyz}
&\phi_{xyz} =  \sqrt{\frac{105}{4\pi}} \frac{xyz}{r^2},\\
\label{eq:xa}
&(\phi_{x\alpha}, \phi_{y\alpha}, \phi_{z\alpha})   \notag\\
&= \frac{1}{2} \sqrt{\frac{7}{4\pi}} \left[\frac{x(5x^2 - 3r^2)}{r^3}, \frac{y(5y^2 - 3r^2)}{r^3}, \frac{z(5z^2 - 3r^2)}{r^3}\right],  \\
\label{eq:xb}
&(\phi_{x\beta},\phi_{y\beta}, \phi_{z\beta}) = \frac{1}{2} \sqrt{\frac{105}{4\pi}} \left[\frac{x(y^2-z^2)}{r^3}, \frac{y(z^2-x^2)}{r^3}, \frac{x(x^2-y^2)}{r^3} \right],
\end{align}
where $r=\sqrt{x^2+y^2+z^2}$.

We also present the expressions of multipoles for $0 \leq l \leq 3$.
By using the real expressions for the multipole operator $\hat{X}_{lm}$, which are given by
\begin{align}
\hat{X}_{lm}^{({\rm c})} &= \frac{1}{\sqrt{2}}\left[\hat{X}_{l-m} + (-1)^m \hat{X}_{lm} \right],\\
\hat{X}_{lm}^{({\rm s})} &= -\frac{1}{\sqrt{2}i}\left[\hat{X}_{l-m} - (-1)^m \hat{X}_{lm} \right],
\end{align}
a monopole, dipoles, quadrupoles, and octupoles are expressed as
\begin{align}
\hat{X}_0 &= \hat{X}_{00},\\
\left(\hat{X}_x, \hat{X}_y, \hat{X}_z \right) &= \left( \hat{X}_{11}^{({\rm c})}, \hat{X}_{11}^{({\rm s})}, \hat{X}_{10} \right),\\
\left( \hat{X}_{u}, \hat{X}_{v}, \hat{X}_{yz},\hat{X}_{zx},\hat{X}_{xy}\right)  &= \left(\hat{X}_{20}, \hat{X}_{22}^{\rm (c)}, \hat{X}_{21}^{\rm (s)},  \hat{X}_{21}^{\rm (c)}, \hat{X}_{22}^{\rm (s)} \right),\\
\hat{X}_{xyz} &= \hat{X}_{32}^{\rm (s)}, \\
\hat{X}_{x\alpha} &= \frac{1}{2\sqrt{2}} \left[\sqrt{5} \hat{X}_{33}^{\rm (c)} - \sqrt{3} \hat{X}_{31}^{\rm (c)}\right], \\
\hat{X}_{y\alpha} &= -\frac{1}{2\sqrt{2}} \left[\sqrt{5} \hat{X}_{33}^{\rm (s)} + \sqrt{3} \hat{X}_{31}^{\rm (s)}\right], \\
\hat{X}_{z\alpha} &= \hat{X}_{30}^{\rm (c)}, \\
\hat{X}_{x\beta} &= -\frac{1}{2\sqrt{2}} \left[\sqrt{3} \hat{X}_{33}^{\rm (c)} + \sqrt{5} \hat{X}_{31}^{\rm (c)}\right], \\
\hat{X}_{y\beta} &= \frac{1}{2\sqrt{2}} \left[-\sqrt{3} \hat{X}_{33}^{\rm (s)} + \sqrt{5} \hat{X}_{31}^{\rm (s)}\right], \\
\hat{X}_{z\beta} &= \hat{X}_{32}^{\rm (c)}.
\end{align}

\section{The Inter- and Intra-Orbital Antisymmetric Spin-Orbit Interaction}
\label{ap:ASOI}
The effective model in Eq.~(\ref{eq:model_Hamiltonian}) implicitly includes the inter-orbital Rashba-type ASOI, as discussed in Sect.~\ref{sec:model} in the main text.
In this Appendix, we discuss the role of the inter-orbital ASOI by taking an appropriate unitary transformation for $(\phi_{d,\sigma}, \phi_{f,\sigma})$.  
As the inter-site hybridization $\mathcal{H}_{\rm mixing}$ in Eq.~(\ref{eq:Hmix}) and the odd-parity CEF $\mathcal{H}_{\rm odd}$ in Eq.~(\ref{eq:Hodd}) are important to derive the inter-orbital ASOI, we only consider these two terms.
Then, the Hamiltonian is explicitly written as
\begin{align}
\label{eq:H_mix_odd}
\mathcal{H} &= \sum_{\bm k} {\bm c}_{\bm k}^\dagger \hat{\mathcal{H}} {\bm c}_{\bm k}, \\
 \hat{\mathcal{H}} &= 
 \left(
 \begin{array}{cccc}
 0 & V_{\rm intra} & 0 &  h({\bm k}) \\
 V_{\rm intra} & 0 &  h({\bm k}) & 0 \\
 0 &  h^*({\bm k})& 0 & V_{\rm intra} \\
 h^*({\bm k}) & 0 & V_{\rm intra} & 0 \\
 \end{array}
 \right),\\
 h({\bm k}) &= 2V_{\rm inter}({\rm sin}k_y + i{\rm sin}k_x),
\end{align}
where ${\bm c}_{\bm k}^\dagger = (d_{{\bm k}\uparrow}^\dagger, f_{{\bm k}\uparrow}^\dagger, d_{{\bm k}\downarrow}^\dagger, f_{{\bm k}\downarrow}^\dagger)$ [${\bm c}_{\bm k} = (d_{{\bm k}\uparrow}, f_{{\bm k}\uparrow}, d_{{\bm k}\downarrow}, f_{{\bm k}\downarrow})$].
By transforming $(\phi_{d,\sigma}, \phi_{f,\sigma})$ to the basis diagonalizing the odd-parity CEF, 
the Hamiltonian in Eq.~(\ref{eq:H_mix_odd}) turns into
\begin{align}
\mathcal{H} &= \sum_{\bm k} {\bm c}_{\bm k}^{\prime\dagger} \hat{{\mathcal{H}}^\prime} {\bm c}^\prime_{\bm k}, \\
\label{eq:H_ASOI_2}
 \hat{{\mathcal{H}}^\prime} &= 
  \left(
 \begin{array}{cccc}
V_{\rm intra} & 0 & h({\bm k}) &  0 \\
0  & -V_{\rm intra} &  0 & -h({\bm k}) \\
 h^*({\bm k}) & 0 & V_{\rm intra} & 0 \\
 0 & -h^*({\bm k})  & 0 & -V_{\rm intra} \\
 \end{array}
 \right),
\end{align} 
where 
 ${\bm c}_{\bm k}^{\prime\dagger} = (a_{{\bm k}\uparrow}^\dagger, b_{{\bm k}\uparrow}^\dagger, a_{{\bm k}\downarrow}^\dagger, b_{{\bm k}\downarrow}^\dagger)$ [${\bm c}^\prime_{\bm k} = (a_{{\bm k}\uparrow}, b_{{\bm k}\uparrow}, a_{{\bm k}\downarrow}, b_{{\bm k}\downarrow})$]. 
$a_{{\bm k}\sigma}^\dagger$ and $b_{{\bm k}\sigma}^\dagger$ ($a_{{\bm k}\sigma}$ and $b_{{\bm k}\sigma}$) are represented by 
 $a_{{\bm k}\sigma}^\dagger = (d_{{\bm k}\sigma}^\dagger+f_{{\bm k}\sigma}^\dagger)/\sqrt{2}$ and $b_{{\bm k}\sigma}^\dagger = (d_{{\bm k}\sigma}^\dagger-f_{{\bm k}\sigma}^\dagger)/\sqrt{2}$ [$a_{{\bm k}\sigma} = (d_{{\bm k}\sigma}+f_{{\bm k}\sigma})/\sqrt{2}$, $b_{{\bm k}\sigma} = (d_{{\bm k}\sigma}-f_{{\bm k}\sigma})/\sqrt{2}$].
 Then, the off-diagonal part in Eq.~(\ref{eq:H_ASOI_2}) represents the inter-orbital ASOI, which is shown as  
\begin{align}
\label{eq:ASOI_inter}
\mathcal{H}_{\rm ASOI} =  2V_{\rm inter}
\sum_{\bm k,\sigma,\sigma^\prime}   
({\rm sin}k_x \sigma_y^{\sigma \sigma^\prime} - {\rm sin}k_y \sigma_x^{\sigma\sigma^\prime})
(b_{{\bm k}\sigma}^\dagger b_{{\bm k}\sigma^\prime} - a_{{\bm k}\sigma}^\dagger a_{{\bm k}\sigma^\prime}).
\end{align}
Note that this type of ASOI is present in a staggered way for orbitals $a$ and $b$.

Next, we discuss the intra-orbital ASOI for $(\phi_{d,\sigma}, \phi_{f,\sigma})$ ignored in the main text.
The intra-orbital ASOI is obtained by considering the higher-order $d$-$f$ hybridization with respect to the odd-parity CEF and the atomic SOC.
\begin{figure}[t!]
\centering
\includegraphics[width=85mm]{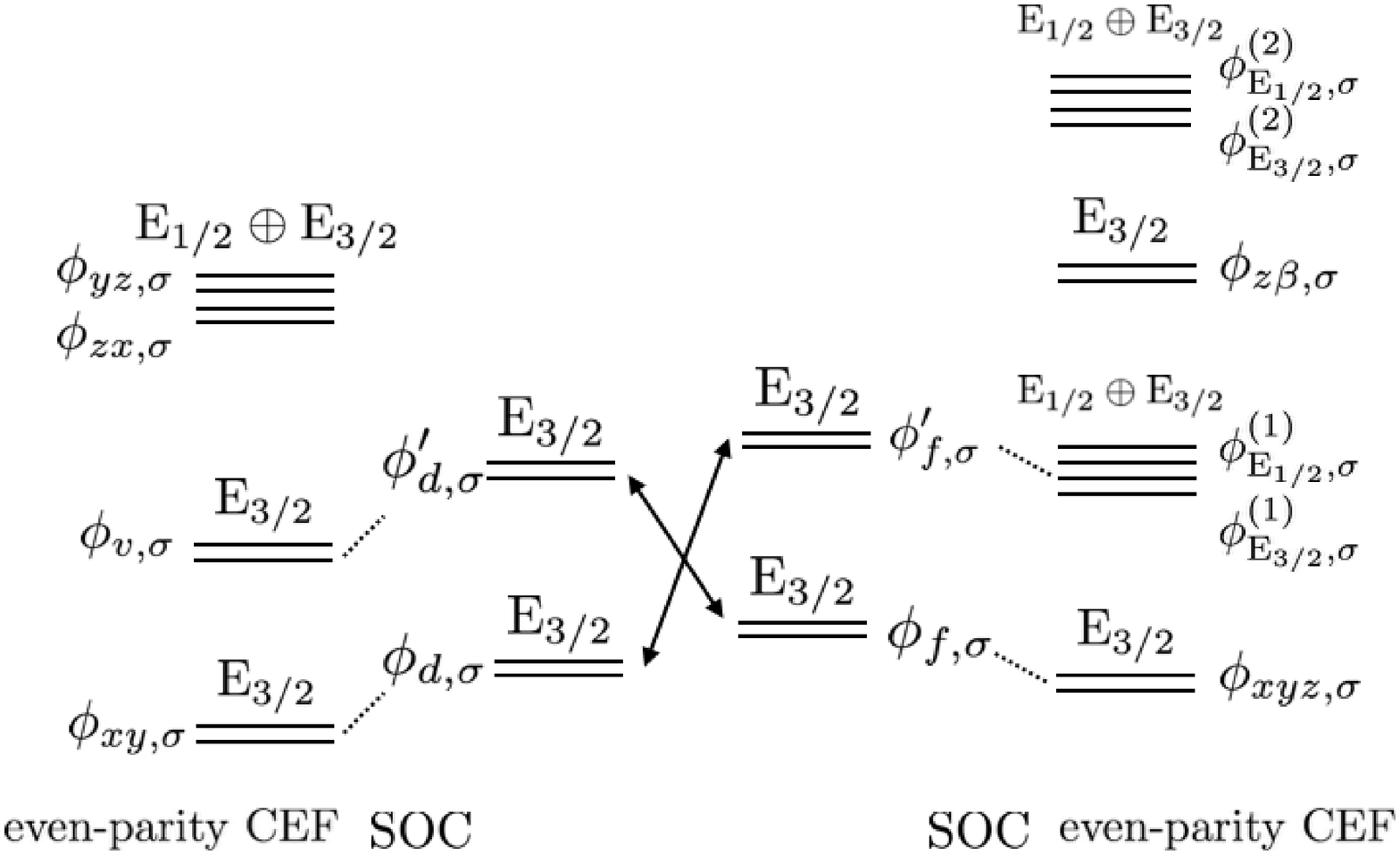}
\caption{The level schemes of the orbitals belongs to the ${\rm E_{3/2}}$ representation. 
$\phi_{d,\sigma}$ and $\phi_{f,\sigma}$ are the $d$ and $f$ orbitals with the lowest energy, and $\phi_{d,\sigma}^\prime$ and $\phi_{f,\sigma}^\prime$ are the $d$ and $f$ orbitals with the second lowest energy .
\label{fig:CEF_split3}}
\end{figure}
Assuming $(\phi_{d,\sigma}^\prime, \phi_{f,\sigma}^\prime)$ has the second lowest energy, 
as shown in Fig.~\ref{fig:CEF_split3} (see also Fig.~\ref{fig:CEF_split} in the main text), 
the wave functions of $(\phi_{d,\sigma}^\prime, \phi_{f,\sigma}^\prime)$ are represented by
\begin{align}
\phi_{d,\sigma}^\prime &\propto  \left[ \phi_{v,\sigma} \pm 2i \Lambda_{\rm B_2}^{\prime d} \phi_{xy,\sigma}- \Lambda_{\rm E}^{\prime d}(i\phi_{yz,\tilde{\sigma}} \mp \phi_{zx, \tilde{\sigma}}) \right], \\
\phi_{f,\sigma}^\prime &\propto  \left[\phi_{{\rm E_{3/2}},\tilde{\sigma}}^{(1)} - \Lambda_{\rm B_2}^{\prime f}i \phi_{xyz,\sigma} \pm \Lambda_{\rm B_1}^{\prime f} \phi_{z\beta, \sigma}
\right],
\end{align}
\begin{align}
\Lambda_{\Gamma}^{\prime d} &= \frac{\lambda_d}{\Delta_{\rm B_1}^{d} - \Delta_{\Gamma}^{d}},\\
\Lambda_{\rm B_2}^{\prime f} &= \frac{2\sqrt{2}\lambda_f}{\Delta_{\rm E}^{(f,1)} - \Delta_{\rm B_2}^{f}}c_1, \\
\Lambda_{\rm B_1}^{\prime f} &= \frac{1}{\sqrt{2}}\frac{\lambda_f}{\Delta_{\rm E}^{(f,1)} - \Delta_{\rm B_1}^{f}}\left(c_1-\sqrt{15}c_2 \right), 
\end{align}
where the subscript $\tilde{\sigma}$ represents the opposite spin component to $\sigma$.
$\Delta_{\Gamma}^{\zeta}$ is the atomic energy level split under the even-parity CEF for $\zeta=d, f$ and $\Gamma = {\rm B_2, E}$. $\Delta_{\rm E}^{(f,1)}$ and $\Delta_{\rm E}^{(f,2)}$ are the energy levels of two pairs of $f$ orbitals belonging to the representation ${\rm E}$.
$\phi_{{\rm E_{3/2}},\sigma}^{(1)}$ is defined in Sect.~\ref{subsec:CEF}.
By considering the odd-parity hybridization between $\phi_{d,\sigma}$ and $\phi_{f,\sigma}^\prime$ ($\phi_{f,\sigma}$ and $\phi_{d,\sigma}^\prime$), the basis functions are modulated as
\begin{align}
\phi_{d,\sigma} &\to
\tilde{\phi}_{d,\sigma} \propto  \left( \phi_{d,\sigma} + \frac{\braket{\phi^\prime_{f,\sigma}|\mathcal{H}_{\rm CEF}^{\rm (odd)}|\phi_{d,\sigma}}}{\Delta_d - \Delta_f^\prime}
\phi_{f,\sigma}^\prime \right),\\
\phi_{f,\sigma} &\to
\tilde{\phi}_{f,\sigma} \propto \left( \phi_{f,\sigma} + \frac{\braket{\phi^\prime_{d,\sigma}|\mathcal{H}_{\rm CEF}^{\rm (odd)}|\phi_{f,\sigma}}}{\Delta_f - \Delta_d^\prime}
\phi_{d,\sigma}^\prime \right),
\end{align}
where $\braket{\phi^\prime_{\zeta^\prime,\sigma}|\mathcal{H}_{\rm CEF}^{\rm (odd)}|\phi_{\zeta,\sigma}}$ for $\zeta, \zeta^\prime= d, f$ ($\zeta \neq \zeta^\prime$) is the order of $V_{\rm intra}\Lambda^{\prime \zeta^\prime}_{\rm B_2}$.

By taking into account the off-site $d$-$f$ hybridization of $(\tilde{\phi}_{d,\sigma},\tilde{\phi}_{f,\sigma})$, the intra-orbital ASOI is obtained as
\begin{align}
\label{eq:intra_ASOI2}
\mathcal{H}_{\rm ASOI}^{\rm intra} =
 \sum_{\bm k} \sum_{\zeta}
\sum_{\sigma,\sigma^\prime}
g_\zeta ({\rm sin}k_x \sigma_y^{\sigma \sigma^\prime} - {\rm sin}k_y \sigma_x^{\sigma\sigma^\prime})
\ \zeta_{{\bm k}\sigma}^\dagger \zeta_{{\bm k}\sigma^\prime} ,
\end{align}
where $g_\zeta$ ($\zeta=d$, $f$) is the magnitude of the intra-orbital ASOI parameter.
The expression in Eq.~(\ref{eq:intra_ASOI2}) corresponds to that in Eq.~(\ref{eq:intra_ASOI1}) in Sect.~\ref{subsec:METz}.
The order of $g_\zeta$ is represented by $g_d \sim V_{\rm intra} V_{df\pi} \Lambda_{\rm B_2}^{\prime f}/(\Delta_d - \Delta_{f}^\prime)$ and $g_f \sim V_{\rm intra} 
V_{df\sigma}  \tilde{\Lambda}_{\rm E}^{(f,i)} \Lambda_{\rm B_2} ^{\prime d} /(\Delta_f - \Delta_{d}^\prime)$ $(i=1,2)$,  
where $V_{df\pi}$ and $V_{df\sigma}$ are the Slater-Koster parameters. 
Thus, the magnitude of the intra-orbital ASOC is smaller than that of the inter-orbital ASOI in Eq.~(\ref{eq:ASOI_inter}) at least by $V_{\rm intra}/(\Delta_{\zeta} - \Delta_{\zeta^\prime}^\prime)$ for $\zeta, \zeta^\prime = d, f$ ($\zeta \neq \zeta^\prime$).

\bibliographystyle{jpsj}
\bibliography{69137}

\begin{thebibliography}{10}
\bibitem{comment_CEF}
{The qualitatively similar results are obtained when the energy scale of the
  SOC is larger than that of the even-parity CEF.}
\end{thebibliography}

\end{document}